\documentclass[a4paper,11pt]{article}
\usepackage{jcappub} 
\usepackage{lineno}
\usepackage{xcolor}   
\usepackage{mathtools}
\usepackage{setspace}

\newcommand{\Om}{\Omega_{\rm m}}
\newcommand{\sig}{\sigma_8}

\newcommand{\Fpar}{F_{\rm{G}}}

\newcommand{\logxicgs}{\log \xi_{\rm cg}^{\rm s}}  

\newcommand{\logxicgr}{\log \xi_{\rm cg}^{\rm r}}  
\newcommand{\xicgs}{\xi_{\rm cg}^{\rm s}} 
\newcommand{\xicgr}{\xi_{\rm cg}^{\rm r}} 
\newcommand{\logxicgsfull}{\log \xi_{\rm cg}^{\rm s}(r_p, r_\pi)} 
\newcommand{\logxicgrfull}{\log \xi_{\rm cg}^{\rm r}(r_p, r_\pi)} 
\newcommand{\xicgsfull}{\xi_{\rm cg}^{\rm s}(r_p, r_\pi)}  
\newcommand{\xicgrfull}{\xi_{\rm cg}^{\rm r}(r_p, r_\pi)} 
\newcommand{\xiggs}{\xi_{\rm gg}^{\rm s}}

\newcommand{\rp}{r_p}
\newcommand{\rpi}{r_\pi}

\newcommand{\vrad}{v_{\rm rad}}
\newcommand{\kms}{\,{\rm km\,s^{-1}}}

\newcommand{\lmcut}{\log M^{\rm cut}_{\rm obs}}
\newcommand{\smscat}{\sigma_{\log M_{\rm c}}}
\newcommand{\siglogM}{\sigma_{\log M}}
\newcommand{\logMmin}{\log M_{\rm min}}

\newcommand{\logMone}{\log M_1}

\newcommand{\alphaHOD}{\alpha}
\newcommand{\etacen}{\eta_{\rm cen}}
\newcommand{\etasat}{\eta_{\rm sat}}
\newcommand{\etasatlow}{\eta_{\rm sat,low}}
\newcommand{\etasathigh}{\eta_{\rm sat,high}}

\newcommand{\DESI}{\textsc{DESI}}

\newcommand{\AbacusSummit}{\textsc{AbacusSummit}}

\newcommand{\Msun}{M_\odot}
\newcommand{\hmpc}{\,h^{-1}{\rm Mpc}}

\defcitealias{Weinberg2024}{WGJT24}

\title{The Redshift-Space Cluster-Galaxy Cross-Correlation as a Probe of Gravity on Cosmic Scales}

\author[a,b,1]{Nicole Marcelina Gountanis\note{Corresponding author.},}
\author[a,b]{David H. Weinberg,}
\author[c,d]{Andr\'{e}s N. Salcedo,}
\author[a,b]{and Claire Lamman}

\affiliation[a]{Department of Astronomy, The Ohio State University,\\
Columbus, OH 43210, USA}
\affiliation[b]{Center for Cosmology and AstroParticle Physics, The Ohio State University,\\
Columbus, OH 43210, USA}
\affiliation[c]{Department of Astronomy/Steward Observatory, University of Arizona,\\
933 North Cherry Avenue, Tucson, AZ 85721-0065, USA}
\affiliation[d]{Department of Physics, University of Arizona,\\
1118 East Fourth Street, Tucson, AZ 85721, USA}

\emailAdd{gountanis.1@osu.edu}

\abstract{Testing General Relativity (GR) on cosmological scales requires probes that are directly sensitive to gravitational dynamics. We present a Fisher matrix forecast for the precision with which the Dark Energy Spectroscopic Instrument (DESI) Bright Galaxy Survey (BGS) can constrain departures from GR through the cluster infall velocity field, using the redshift-space cluster--galaxy cross-correlation function $\xicgsfull$ as the primary observable. We construct mock galaxy and cluster catalogs from the \textsc{AbacusSummit} $N$-body simulations using a Halo Occupation Distribution (HOD) framework for galaxies and a log-normal mass--observable relation for cluster selection. We parameterize departures from GR with a single multiplicative factor $\Fpar$ that rescales halo infall velocities relative to the GR prediction. Marginalizing simultaneously over the HOD, mass--observable, and cosmological parameters, we find that DESI BGS can constrain $\Fpar$ to $\sigma(\Fpar) = 0.033$ at $z \sim 0.3$. This precision is sufficient to distinguish representative $f(R)$ and Galileon gravity models from GR at $\gtrsim 4\sigma$. The strongest limitation is a degeneracy between $\Fpar$ and the scatter in the cluster mass--observable relation, while the constraint shows no strong degeneracy with any individual HOD parameter. Perfect knowledge of the galaxy--halo connection would improve the constraint to $\sigma(\Fpar) \approx 0.015$, and including the galaxy autocorrelation function in the data vector could recover part of this improvement. The same data vector yields tight constraints on the cosmological parameters, $\sigma(\Omega_m) = 0.010$ and $\sigma(\sigma_8) = 0.010$, with only moderate correlation between these parameters and $\Fpar$. These results demonstrate the potential of DESI BGS cluster--galaxy redshift-space correlations to test gravitational physics on quasi-linear scales, complementary to existing probes of modified gravity and to weak-lensing measurements of cluster mass profiles.}

\begin{document}
\maketitle
\flushbottom

\section{Introduction}\label{sec:intro}

After three decades of intensive study, the origin of cosmic acceleration remains the most profound mystery of contemporary cosmology. Does accelerating expansion signal a breakdown of General Relativity (GR) on cosmological scales, or is it caused by an exotic form of energy that exerts repulsive gravity within GR? The example of a cosmological constant shows that the answer to this question may be ambiguous, as $\Lambda g_{\mu\nu}$ can be added to the curvature or stress-energy side of the GR field equation \citep{Einstein1917, Zeldovich1968}. However, many modifications of GR give rise to distinctive gravitational effects in addition to changing the expansion equation, such as growth of structure at a rate that depends on the redshift, scale, or environment (see reviews in refs.~\cite{Jain2010, Baker2019}). Furthermore, the bending of light and acceleration of non-relativistic particles depend on distinct gravitational potentials that are equal at late times in GR but may differ in modified gravity theories, so that mass estimates based on gravitational lensing and peculiar velocities can yield incompatible results \citep{Zhang2007, Reyes2010}. Building on previous studies by Zu \& Weinberg~\cite{zuw2013} and Zu et al.~\cite{Zu2014}, we investigate the cluster-galaxy redshift-space cross-correlation function $\xicgsfull$ as a probe of modified gravity and a test of GR.

Many modified gravity theories take the form of tensor-scalar models, in which a scalar field sources an acceleration acting as a ``fifth force'' alongside the standard gravitational interaction of the tensor field \citep[e.g.,][]{Horndeski1974, HuSawicki2007, Jain2010, Baker2019}. Since GR makes extremely accurate predictions on solar system and stellar scales, including binary pulsars and merging black holes, any scalar field contribution must be strongly suppressed in these environments. In empirically viable theories, this suppression emerges naturally from non-linear scalar field interactions, giving rise to screening mechanisms that operate in regions of high potential or high density \citep{vainshtein1972, khouryweltman2004}. As the universe evolves, screening is effective at early times but weakens in regions near the cosmic mean density at late times, where gravitational accelerations can depart from GR predictions at order unity and produce substantial deviations in galaxy peculiar velocities. In $f(R)$ gravity, for instance, where the Ricci scalar $R$ in the Einstein-Hilbert action is replaced by $R + f(R)$, accelerations in unscreened regions are enhanced by a factor of $4/3$ \citep{Burrage2018}. The same enhancement appears in the unscreened, static regime of the normal branch of the DGP model \citep{Dvali2000, Koyama2007}, where modification of gravity on cosmic scales arises from the presence of an additional spatial dimension.

The infall regions around massive galaxy clusters offer a novel opportunity to test the GR-predicted velocity scale. Within the virial radius, the velocity dispersion of member galaxies produces the well-known Finger-of-God (FoG) distortion, stretching clusters along the line of sight in redshift space. At larger distances, coherent infall produces a squashing distortion, including caustics near the turnaround radius, where infall velocities cancel the Hubble flow \citep{Sargent1977, Kaiser1987, Regos1989, Diaferio1997, Gifford2013}. Zu \& Weinberg~\cite{zuw2013} developed an analytic model for $\xicgsfull$, and used it to extract the characteristic radial velocity profile $v_{\rm c}(r)$ as a function of the three-dimensional radius, $r=(\rp^2 + \rpi^2)^{1/2}$. Applying this model to group catalogs \citep{Yang2007} from the Sloan Digital Sky Survey \citep[SDSS;][]{York2000, Strauss2002}, they recovered $v_{\rm c}(r)$ profiles in good agreement with predictions from the \textsc{Millennium} $N$-body simulation \citep{Springel2005}, with maximum infall velocities of $600$--$800\ \rm km\,s^{-1}$.

Zu et al.~\cite{Zu2014} applied the same analysis to $N$-body simulations of $f(R)$ gravity and a Galileon model, which follow the chameleon \citep{khouryweltman2004} and Vainshtein \citep{vainshtein1972} screening mechanisms, respectively. In both cases, the characteristic infall velocities $v_{\rm c}(r)$ were enhanced by $\sim 50$--$150\ \rm km\,s^{-1}$ relative to the GR predictions for cluster populations of equal space density, attributable to the higher gravitational accelerations in unscreened environments at large distances from clusters. However, these enhancements of $10$--$30\%$ were comparable in magnitude to the uncertainties in the SDSS analysis of Zu \& Weinberg~\cite{zuw2013}, preventing a definitive test of the modified gravity models.

This work revisits the analysis of Zu \& Weinberg~\cite{zuw2013} and Zu et al.~\cite{Zu2014} for two reasons. First, the Dark Energy Spectroscopic Instrument \citep[DESI;][]{DESICollaboration2016} Bright Galaxy Survey \citep[BGS;][]{hahn2023} provides a galaxy redshift survey ten times larger than the SDSS main galaxy survey at a similar number density \cite{Levi2013}, substantially reducing the statistical uncertainties that limited the earlier analysis. Second, computational advances now make it feasible to model $\xicgsfull$ directly within a parameter inference framework, bypassing the need for the analytic model of Zu \& Weinberg~\cite{zuw2013}. Recent studies of galaxy-galaxy lensing, galaxy redshift-space correlations, and cluster weak lensing have increasingly relied on simulation-based halo modeling, using emulators \citep{Wibking2020, Salcedo2022, Zhai2023, Yuan2022a, Yuan2022b, Paillas2024}, subhalo abundance matching \citep{Contreras2023a, Contreras2023b, Ortega-Martinez2026}, or direct halo occupation modeling of simulations \citep{Lange2025, Lange2023, Lange2022}. We plan to construct an emulator for $\xicgsfull$ in future work. For this paper, we evaluate derivatives of observables from perturbations around a fiducial model and use these to carry out a Fisher matrix forecast, illustrating the general principles of our approach and the sensitivity achievable with DESI BGS.

In this work, we use GR simulations from the \textsc{AbacusSummit} $N$-body suite \citep{garrison21, maksimova21}, adopting a simple parameterization to capture modified gravity effects by scaling all halo velocities by a factor $\Fpar$, where $\Fpar = 1$ recovers the standard GR prediction. This choice is motivated by the linear growth rate parameter $f(\Om) \approx \Om^{0.55}$, which scales peculiar velocities relative to displacements in linear perturbation theory. Following the results of Zu et al.~\cite{Zu2014}, applying this methodology to modified gravity simulations is expected to yield best-fit values of $\Fpar > 1$, with $1.1$--$1.3$ typical for the models examined therein. We defer analysis of modified gravity simulations to future work. While our parameterization does not map onto a specific modified gravity theory, our forecast addresses a well-defined question: with what precision can a survey like the DESI BGS test the GR-predicted scale of peculiar velocities in cluster infall regions?

We focus on the $z = 0.3$ snapshots of the \textsc{AbacusSummit} simulations and present forecasts scaled to the DESI DR3 BGS volume in the redshift range $z = 0.25$--$0.35$. Redshift evolution across $z$-bins and the potential gains from analysing samples at lower and higher redshifts are left to future work. We adopt a conventional but idealized cluster selection model, in which each observed cluster is drawn from a population of massive halos whose observational mass estimates are related to their true three-dimensional masses by log-normal scatter. Anisotropic selection biases are known to affect predicted weak lensing profiles significantly \citep{Wu2022, Sunayama2020}. Analogous biases are expected to be important for $\xicgsfull$, even if clusters are identified from spectroscopic galaxy catalogs, since galaxies within $\sim 1000\ \rm km\,s^{-1}$ velocity separations can reside within the cluster or in the foreground or background \citep{Wu2022, Lee2025, Myles2025}. Experience from cluster weak lensing analyses demonstrates that full forward modeling of cluster selection within an $N$-body simulation framework can mitigate or eliminate such biases, enabling robust cosmological inference \citep{Salcedo2024, Salcedo2025}. We anticipate that a similar framework will be required for accurate modeling of $\xicgsfull$ with DESI BGS, but we defer its development to future work.

With the high precision potentially achievable by DESI BGS, cosmological constraints from $\xicgsfull$ can complement those from cluster weak lensing in large photometric surveys (e.g., refs.~\cite{Rozo2010, VonDerLinden2014, Zu2014b, To2021, Abbott2025, Salcedo2024, Salcedo2025}). Like stacked weak lensing profiles, redshift-space distortions of $\xicgsfull$ probe the virial masses and extended mass profiles of cluster halos. In combination with cluster number counts, these measurements constrain the mean matter density $\Omega_m$ and the amplitude of the matter power spectrum $\sigma_8$ at the typical cluster mass scale \citep{White1993}. We give preliminary estimates of the sensitivity to $\sigma_8$ and $\Omega_m$ in our Fisher forecasts, and we find only moderate correlation between the constraints on these parameters and those on $\Fpar$.

Ultimately, we aim to combine $\xicgsfull$ with weak lensing to test the consistency of the gravitational potentials that govern light bending and galaxy motions. Light bending and galaxy motions respond differently to the two scalar potentials that describe the space and time components of the perturbed Friedmann-Robertson-Walker metric (see, e.g., ref.~\cite{Zhang2007}). These potentials are equal in magnitude in GR\footnote{The two potentials can differ in the presence of anisotropic stress, though this effect is negligible at late times in a cosmological model with standard matter and energy content.} but can differ in modified gravity models. The $E_{\rm{G}}$ statistic, also known as the gravitational slip test, combines galaxy-galaxy lensing and galaxy redshift-space distortions to test the consistency of these potentials \citep{Zhang2007, Reyes2010, Rauhut2025}. Our $\Fpar$ parameterization is motivated partly by analogy to $E_{\rm{G}}$, and future joint analyses of $\xicgsfull$ and cluster weak lensing could carry out an analogous test in the physically distinct environment of cluster infall regions.

The remainder of this paper is organized as follows. Section~\ref{sec:sims} describes the \AbacusSummit\ simulations and mock galaxy and cluster catalogs. Section~\ref{sec:infall} characterizes the mean radial infall velocity profile and its dependence on each model parameter. Section~\ref{sec:xicgs} introduces $\xicgsfull$ and our measurement methodology. Section~\ref{sec:sensitivity} quantifies the sensitivity of the full data vector to each parameter through derivatives of $\logxicgs$, $\log n_{\rm gal}$, and $\log n_{\rm halo}$. Section~\ref{sec:fisher} presents the Fisher matrix formalism and the Full Forecast constraints. Section~\ref{sec:discussion} examines the information content of the data vector, the role of the infall velocity field in driving the constraints, and the steps required to apply this methodology to DESI BGS data. Section~\ref{sec:conclusions} summarizes our main conclusions.

\section{Simulations and Mock Catalogs} \label{sec:sims}

We construct our mock galaxy and cluster catalogs from the \AbacusSummit\ $N$-body simulation suite, which provides the large-scale dark matter distribution needed to model the cluster infall regime. To populate these simulations, we model the galaxy--halo connection using a Halo Occupation Distribution (HOD) framework. We also account for the stochastic mapping between observable cluster proxies and true halo mass through a log-normal mass--observable relation. From these catalogs, we compute the clustering ($\log\xicgs$) and number density ($\log n_{\rm gal}$ and $\log n_{\rm halo}$) observables that form our data vector.

\subsection{AbacusSummit Simulations}
\label{sec:abacus}

\begin{table}
    \centering
    \caption{\AbacusSummit\ boxes used in this work. Boxes c102 and c103 vary $\Omega_m$ at fixed $\sigma_8$, while c112 and c113 vary $\sigma_8$ at fixed $\Omega_m$, relative to the fiducial \textit{Planck} cosmology. $\Omega_m$ and $\sigma_8$ values are quoted at $z=0$.}
    \label{tab:cosmo}
    \begin{tabular}{llcccc}
        \hline\hline
        Box & Description & $\Om$ & $A_s\,[10^{-9}]$
        & $\sig$ & $h$ \\
        \hline
        c000 & \textit{Planck} & $0.3152$ & $2.0830$
        & $0.8080$ & $0.6736$ \\
        c102 & high $\Om$ & $0.3378$ & $2.0205$
        & $0.8083$ & $0.6597$ \\
        c103 & low $\Om$ & $0.2942$ & $2.1541$
        & $0.8081$ & $0.6877$ \\
        c112 & high $\sig$ & $0.3152$ & $2.1672$
        & $0.8241$ & $0.6736$ \\
        c113 & low $\sig$ & $0.3152$ & $2.0021$
        & $0.7921$ & $0.6736$ \\
        \hline
    \end{tabular}
\end{table}

\AbacusSummit\ \citep{garrison21, maksimova21} is a suite of large volume, high-accuracy cosmological $N$-body simulations designed to meet the strict modeling requirements of \DESI\ \citep{desicollaboration2024}. The standard boxes contain $6912^3$ particles in a $(2\ h^{-1}\ \mathrm{Gpc})^3$ volume, corresponding to a particle mass resolution of $m_p \approx 2 \times 10^9\ h^{-1}\ M_\odot$, sufficient to resolve the $M \gtrsim 10^{12}\ h^{-1}\ M_\odot$ halos that host typical bright galaxies, which serve as tracers of cluster infall regions.

We adopt the \textit{Planck} cosmology \citep{plackcollaboration2020} as our fiducial cosmology. We use the 25 independent phase realizations of the fiducial cosmology (c000) to estimate the covariance of $\log\xicgs$, $\log n_{\rm gal}$, and $\log n_{\rm halo}$, and their derivatives with respect to the HOD, modified gravity, and mass--observable parameters. We compute derivatives with respect to the cosmological parameters using paired simulation boxes that vary each parameter above and below the fiducial value: c102 and c103 for $\Omega_m$, and c112 and c113 for $\sigma_8$. Table~\ref{tab:cosmo} lists the parameter configurations for each box, including $\Omega_m$, the primordial scalar amplitude $A_s$, $\sigma_8$, and the dimensionless Hubble parameter $h$.\footnote{The complete list of cosmological parameters for all \AbacusSummit\ models is available at \url{https://abacussummit.readthedocs.io/en/latest/cosmologies.html}} In the \AbacusSummit\ cosmology grid, $h$ is not an independent parameter. Instead, $H_0$ is chosen in each box so that the angular size of the sound horizon at last scattering, $\theta_*$, matches the value measured by \textit{Planck}. Therefore, the values of $h$ listed for c102 and c103 shift along with $\Omega_m$, rather than varying independently. 

In this paper, we focus on the simulation snapshots at $z = 0.3$. To forecast constraints from DESI BGS (Section~\ref{sec:results}), we scale the covariance matrix to a comoving volume $V_{\rm survey} = 0.76\ h^{-3}\ {\rm Gpc}^3$, corresponding to the redshift interval $z = 0.25$--$0.35$ over a sky area of $14{,}000\ {\rm deg}^2$. We choose this interval because DESI BGS maintains a high galaxy number density $\bar{n} > 3 \times 10^{-3}\ h^3\ {\rm Mpc}^{-3}$ out to $z = 0.35$ \citep{hahn2023}, roughly corresponding to a volume-limited sample with $M_r \lesssim -20.5$ in the DESI Y3 BGS-BRIGHT sample \citep{FernandezGarcia2025}. The high number density reduces shot noise in $\xicgsfull$ measurements, and it would allow a split into red and blue galaxy subsamples to test whether modified gravity signals and other parameter constraints are insensitive to the modeling of galaxy bias and the galaxy--halo connection. 

We assume that the evolution of structure and HOD parameters can be neglected over a $\Delta z = 0.1$ interval, though this assumption will require verification in future work. DESI BGS would also enable measurements in the redshift intervals $0.15 < z < 0.25$ and $0.35 < z < 0.45$, with comoving volumes approximately $0.5\times$ and $1.6\times$ that of the fiducial $0.25 < z < 0.35$ interval, respectively. In the higher-redshift bin, it may be preferable to use the faint BGS sample, which offers a higher number density at the cost of more complex selection and completeness. At still higher redshifts, applying our method to luminous red galaxy (LRG) samples may be feasible, though their substantially lower number density and higher galaxy bias will require dedicated forecasts to assess sensitivity. 

\subsection{Mock Cluster Catalogs}\label{sec:mock_halos}

Cluster samples are selected via observable mass proxies such as X-ray luminosity, the Sunyaev--Zel'dovich signal \citep{Sunyaev1970}, or optical richness, rather than the underlying true halo mass. Each of these proxies only approximately traces the true halo mass \citep{Kravtsov2006, Evrard2008}. Halos of the same true mass can differ in formation history, structure, and environment, so even a noiselessly measured proxy retains an intrinsic scatter with respect to the true mass \citep{Kravtsov2006, Evrard2008}. Observational measurement uncertainty adds further scatter to the mass-observable relation. Intrinsic and observational scatter shift halos across any fixed selection threshold on the mass proxy, scattering lower-mass systems into the sample and higher-mass systems out of it. Neglecting this effect would introduce a significant bias into our forecast, as the resulting cluster sample would not reflect the realistic mass distribution of an observed cluster catalog.
 
To construct a physically realistic cluster catalog, we begin with the halo catalogs identified by the \textsc{CompaSO} halo finder \citep{hadzhiyska21}. We assign each halo an observed mass proxy $M_{\rm obs}$ by drawing from a log-normal distribution centered on the true halo mass $M_{\rm true}$,
\begin{equation}
\log M_{\rm obs} = \log M_{\rm true} + \epsilon, \quad \epsilon \sim \mathcal{N}(0, \smscat),
\label{eq:mass_scatter}
\end{equation}
where $\smscat$ denotes the combined logarithmic scatter (intrinsic plus measurement) between the proxy and the true cluster mass, in dex. We then define the cluster sample by selecting halos with $\log M_{\rm obs} \geq \lmcut$. This selection allows halos with true masses below the threshold to scatter into the sample and halos above the threshold to scatter out. 
 
\begin{figure}[t]
\centering
\includegraphics[width = \textwidth]{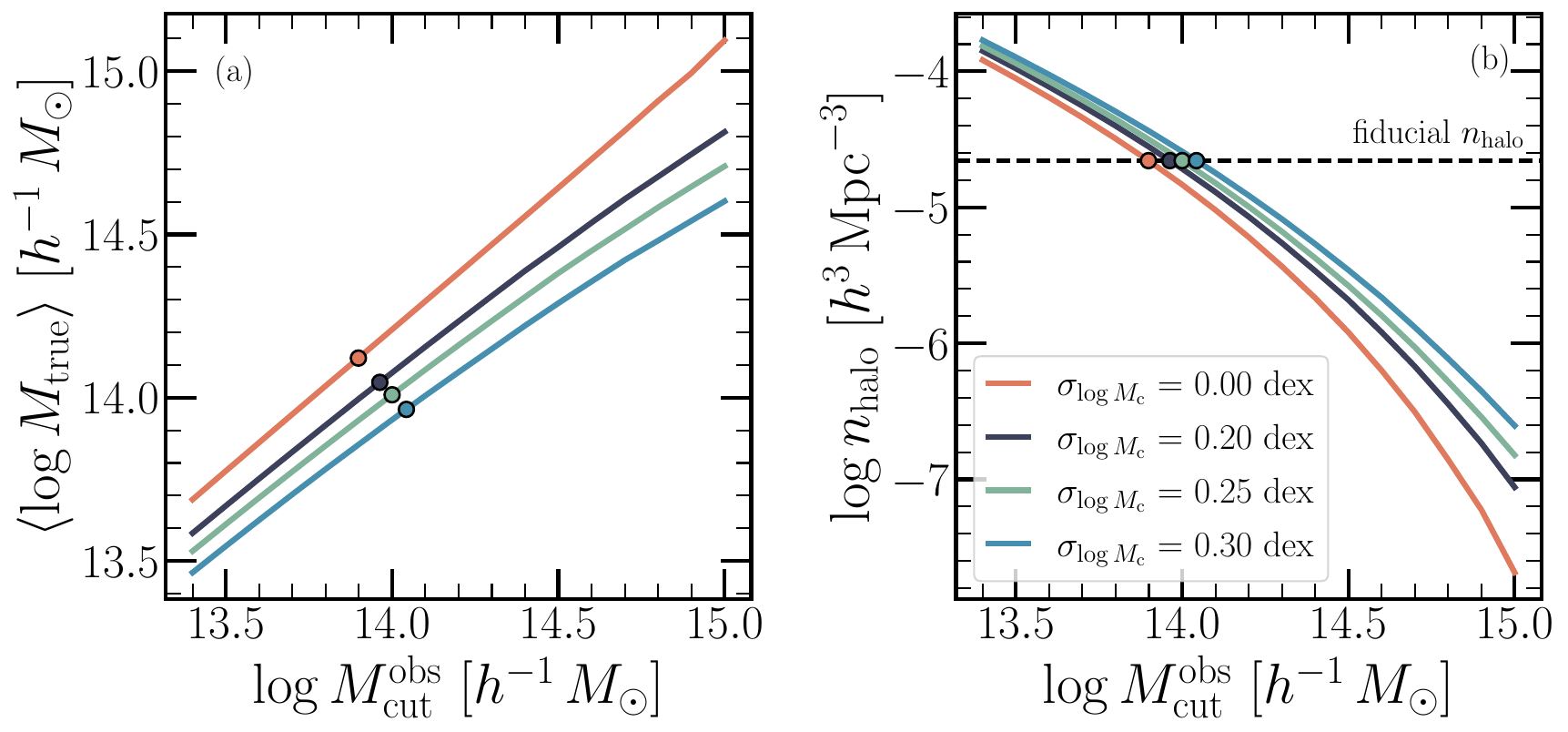}
\vspace{-5mm}
\caption{Mean true halo mass $\langle \log M_{\rm true} \rangle$ (panel~(a)) and halo number density $\log n_{\rm halo}$ (panel~(b)) as a function of the observable mass threshold $\lmcut$ for realization \texttt{ph000} of our fiducial cosmology. Curves show four values of the log-normal scatter: $\smscat = 0$ (red), $0.2$ (navy), $0.25$ (green), and $0.3$~dex (blue). The horizontal dashed line in panel~(b) marks the halo number density of our fiducial model ($\lmcut = 14.0$, $\smscat = 0.25$~dex), $n_{\rm halo} = 2.20 \times 10^{-5}\,h^3\,{\rm Mpc}^{-3}$ ($\log n_{\rm halo} = -4.66$). For each value of $\smscat$, points mark the threshold that yields this number density (panel~(b)) and the corresponding mean true mass (panel~(a)). At the fiducial number density, increasing the scatter from 0 to 0.3~dex lowers the mean true mass of the sample by $\approx 0.16$~dex.}
\label{fig:mcut}
\end{figure}

We treat both $\lmcut$ and $\smscat$ as free parameters in our forecast, so the resulting halo number density $\log n_{\rm halo}$ varies across the parameter space. We include $\log n_{\rm halo}$ as an observable in the data vector. Figure~\ref{fig:mcut} illustrates how variations in $\lmcut$ and $\smscat$ alter the mean mass and comoving number density of the cluster sample.

In each panel, the red curve shows the limiting case of zero scatter ($\smscat=0$), which gives the highest mean mass $\langle \log M_{\rm true} \rangle$ and the lowest cumulative number density $\log n_{\rm halo}$ of clusters above the observable threshold $\lmcut$. Increasing $\smscat$ allows lower-mass halos to enter the sample, increasing $\log n_{\rm halo}$ while reducing $\langle \log M_{\rm true} \rangle$. For our fiducial model parameters, $\lmcut = 14.0$ and $\smscat = 0.25$~dex, the number density is $n_{\rm halo} = 2.20 \times 10^{-5}\,h^3\,{\rm Mpc}^{-3}$, marked by the horizontal dashed line in panel~(b). Points along the curves in panel~(a) show $\langle \log M_{\rm true} \rangle$ at this number density for each value of $\smscat$. At the fiducial number density, increasing the scatter from 0 to 0.3~dex lowers the mean true mass by $\approx 0.16$~dex.
 
\subsection{Mock Galaxy Catalogs} \label{sec:mockgal}

We construct galaxy catalogs by populating the \textsc{CompaSO} halos using a Halo Occupation Distribution (HOD) framework\footnote{\url{https://github.com/ANSalcedo/cHOD}} \citep[e.g.,][]{jing1998J, ma2000, peacock2000, seljak2000, scoccimarro2001, berlind2002, cooray2002}, which specifies the statistical distribution of galaxies within halos as a function of halo mass. The HOD formalism allows us to marginalize over the uncertain physics governing the galaxy--halo connection when deriving constraints on $\Fpar$ or on cosmological parameters ($\sigma_8$, $\Omega_{\rm m}$). We include the galaxy number density $\log n_{\rm gal}$ as an observable in the data vector alongside $\log \xicgs$ and $\log n_{\rm halo}$, providing additional leverage to constrain the HOD parameters. In principle, the galaxy autocorrelation function could provide further HOD constraints \citep[e.g.][]{Zehavi2005, zehavi2011}, but we do not include such constraints here.

The mean occupation numbers for central and satellite galaxies are given by
\begin{align}
    \langle N_{\rm cen}(M) \rangle &= \frac{1}{2}
    \left[1 + \mathrm{erf}\!\left(\frac{\log M -
    \logMmin}{\siglogM}\right)\right],
    \label{eq:Ncen} \\
    \langle N_{\rm sat}(M) \rangle &= \langle
    N_{\rm cen}(M) \rangle \left(\frac{M - M_0}
    {M_1}\right)^{\alphaHOD},
    \label{eq:Nsat}
\end{align}
following ref.~\cite{zheng2005}. The parameter $\logMmin$ sets the characteristic halo mass at which the central occupation reaches $0.5$, and $\siglogM$ controls the sharpness of this transition. The satellite occupation follows a power law with slope $\alphaHOD$ above a low-mass cutoff $M_0$, normalized such that $\langle N_{\rm sat}(M_0 + M_1)\rangle = 1$, where $M_1$ is the characteristic satellite mass scale parametrized by $\logMone \equiv \log_{10}(M_1 / h^{-1}\,\Msun)$. We fix $\log_{10}(M_0/h^{-1}\,\Msun) = 10.0$, well below the minimum resolved halo mass, so the satellite occupation is effectively a pure power law, $\langle N_{\rm sat}\rangle \approx \langle N_{\rm cen}\rangle (M/M_1)^{\alphaHOD}$. We draw the number of centrals per halo from a Bernoulli distribution with mean $\langle N_{\rm cen}(M)\rangle$, and the number of satellites from a Poisson distribution with mean $\langle N_{\rm sat}(M)\rangle$.

We place central galaxies at the center of their host halo, and distribute satellite galaxies spatially according to a Navarro--Frenk--White profile \citep[NFW;][]{nfw1997},
\begin{equation}
    \rho_{\rm gal}(r) = \rho_m(r \mid c_{\rm vir}),
\end{equation}
where $c_{\rm vir} = r_h/r_s$ is the halo concentration. We assign halo concentrations using the mass--concentration relation of ref.~\cite{correa2015}, rescaled from $M_{200c}$ to $M_{200b}$ by multiplying by $\sqrt{2}$ \citep{hu2003}.

Galaxies in spectroscopic surveys are observed in redshift space, where peculiar velocities displace their apparent line-of-sight positions. In our calculations, we treat each of the three Cartesian axes of the simulation cube in turn as the line-of-sight direction and average the results (section~\ref{sec:xicgs}). Since baryons and dark matter experience the same gravitational acceleration, the mean velocity of galaxies in a halo should on average match that of the halo itself. However, the velocity dispersion of satellite galaxies within a halo can differ from that of the dark matter, and the central galaxy may have a non-zero velocity relative to the halo center of mass. Following standard practice for redshift-space galaxy clustering, we characterize these differences through velocity bias parameters \citep{Guo2015}. For a halo with three-dimensional dark matter velocity dispersion $\sigma_v$, we assign each Cartesian component of the central galaxy velocity a stochastic offset drawn from a Laplace distribution with scale parameter $\etacen\,\sigma_v/\sqrt{6}$, where a factor of $\sqrt{3}$ converts $\sigma_v$ to the one-dimensional dispersion and a factor of $\sqrt{2}$ converts the one-dimensional dispersion to the Laplace scale parameter. Similarly, we draw each Cartesian component of the satellite velocity relative to the halo mean velocity from a Gaussian with dispersion $\etasat\,\sigma_v/\sqrt{3}$, so that $\etasat = 1$ corresponds to satellites tracing the velocity dispersion of the dark matter. Velocity bias may depend on the halo mass, and the satellite kinematics of infalling group-mass halos may differ systematically from those of massive cluster-mass halos. We therefore introduce separate parameters $\etasatlow$ and $\etasathigh$ that describe the satellite velocity bias in halos below and above $10^{13.5}\ h^{-1}\ M_\odot$, respectively.

We list the fiducial values for all varied parameters in Table~\ref{tab:params}, with cosmological parameter configurations given separately in Table~\ref{tab:cosmo}. At $z = 0.3$, our fiducial HOD choices give a comoving galaxy space density $n_{\rm gal} = 2.76 \times 10^{-3}\ h^3\ {\rm Mpc}^{-3}$ ($\log n_{\rm gal} = -2.559$), averaged over 25 simulation phases, comparable to the density of BGS-BRIGHT galaxies at this redshift \cite{hahn2023}. The large-scale galaxy bias factor computed from $(\xi_{\rm gg}/\xi_{\rm mm})^{1/2}$ on scales $20$--$30\,h^{-1}\,{\rm Mpc}$, with $\xi_{\rm mm}$ the linear-theory matter correlation function of our fiducial cosmology at $z = 0.3$, is $b_g \approx 1.4$.

\begin{table}
    \centering
    \small
    \caption{Fiducial model parameters and finite-difference step sizes $\delta\theta$ used to compute the derivatives for the Fisher forecast. Cosmological parameter configurations are listed separately in table~\ref{tab:cosmo}.}
    \label{tab:params}
    \begin{tabular}{llcc}
        \hline\hline
        Parameter & Definition & Fiducial & $\delta\theta$ \\
        \hline
        $\siglogM$ & Log-normal scatter width of central occupation transition & $0.60$ & $0.10$ \\
        $\logMmin$ & $\log_{10}$ of characteristic minimum halo mass $[h^{-1}\,\Msun]$ & $12.50$ & $0.10$ \\
        $\logMone$ & $\log_{10}$ of satellite mass scale $M_1$ $[h^{-1}\,\Msun]$ & $13.35$ & $0.10$ \\
        $\alphaHOD$ & Power-law index of satellite occupation & $1.00$ & $0.10$ \\
        $\etacen$ & Central-galaxy velocity bias & $0.10$ & $0.10$ \\
        $\etasatlow$ & Satellite velocity bias for $M < 10^{13.5}\,h^{-1}\,\Msun$ & $1.00$ & $0.10$ \\
        $\etasathigh$ & Satellite velocity bias for $M \geq 10^{13.5}\,h^{-1}\,\Msun$ & $1.00$ & $0.10$ \\
        $\smscat$ & Scatter in mass--observable relation [dex] & $0.25$ & $0.05$ \\
        $\lmcut$ & $\log_{10}$ of observable mass threshold $[h^{-1}\,\Msun]$ & $14.00$ & $0.25$ \\
        $\Fpar$ & Phenomenological modified gravity parameter & $1.00$ & $0.33$ \\
        \hline
    \end{tabular}
\end{table}

\section{Infall Velocities}\label{sec:infall}

The mean radial peculiar velocity profile $\vrad(r)$ of galaxies around massive clusters is the central physical quantity of this analysis. It encodes the coherent kinematic response of matter to the cluster gravitational potential and, as discussed in section~\ref{sec:intro}, is directly sensitive to gravitational accelerations in the quasi-linear infall regime. Before connecting $\vrad(r)$ to the observable redshift-space cross-correlation function $\xicgs$ in section~\ref{sec:xicgs}, we first characterize how it responds to variations in each model parameter.

\subsection{Physical Motivation and Parameterization}\label{sec:phenom}

Galaxy clusters are the most massive gravitationally bound structures in the Universe, and their deep potential wells drive coherent infall of surrounding matter out to large distances. In many modified gravity theories, a scalar field contributes to the gravitational acceleration alongside the metric, enhancing infall velocities relative to the GR prediction at fixed cluster mass. Acceleration is enhanced in unscreened regions, while GR is restored in screened regions; for some broad classes of models, the enhancement in unscreened regions is a factor of 4/3 \citep{Burrage2018}. Since galaxy velocities accumulate over time, they may remain enhanced relative to GR even in regions that have since been screened \citep{Zu2014}.

Rather than simulating specific modified gravity models, we adopt a phenomenological approach in which departures from GR are parameterized by a single multiplicative factor $\Fpar$ that uniformly rescales the mean velocities of halos relative to the GR prediction encoded in the \AbacusSummit\ simulations. This factor consequently rescales mean infall velocities,
\begin{equation}
\vrad(r) \rightarrow \Fpar \, \vrad^{\rm GR}(r),
\label{eq:fpar}
\end{equation}
where $\vrad^{\rm GR}(r)$ denotes the baseline velocity profile measured in our GR simulation suite. In modified gravity models we generally expect $1 \leq \Fpar \leq 4/3$, though our parameterization permits $\Fpar < 1$. Zu et al.~\cite{Zu2014} find that $f(R)$ and Galileon gravity produce qualitatively similar enhancements in the cluster infall velocity profile relative to GR, despite their distinct screening mechanisms. At $r = 5\ h^{-1}\ \mathrm{Mpc}$, the infall velocity enhancement in the $f(R)$ model is approximately $30\%$ for $f_0 = 10^{-4}$ and $15\%$ for $f_0 = 10^{-5}$ (see their figure~5). Comparable but slightly smaller enhancements are seen in their Galileon models. A uniform velocity rescaling captures the dominant signature of modified gravity in a model-independent manner. In future work, we plan to extend this parameterization to allow for scale- or mass-dependent $\Fpar$ calibrated directly against modified gravity simulations.

\subsection{Dependence of the Infall Velocity Profile on Model Parameters}
\label{sec:vrad_dependence}

\begin{figure*}[t]
\centering
\includegraphics[width=0.9\textwidth]
{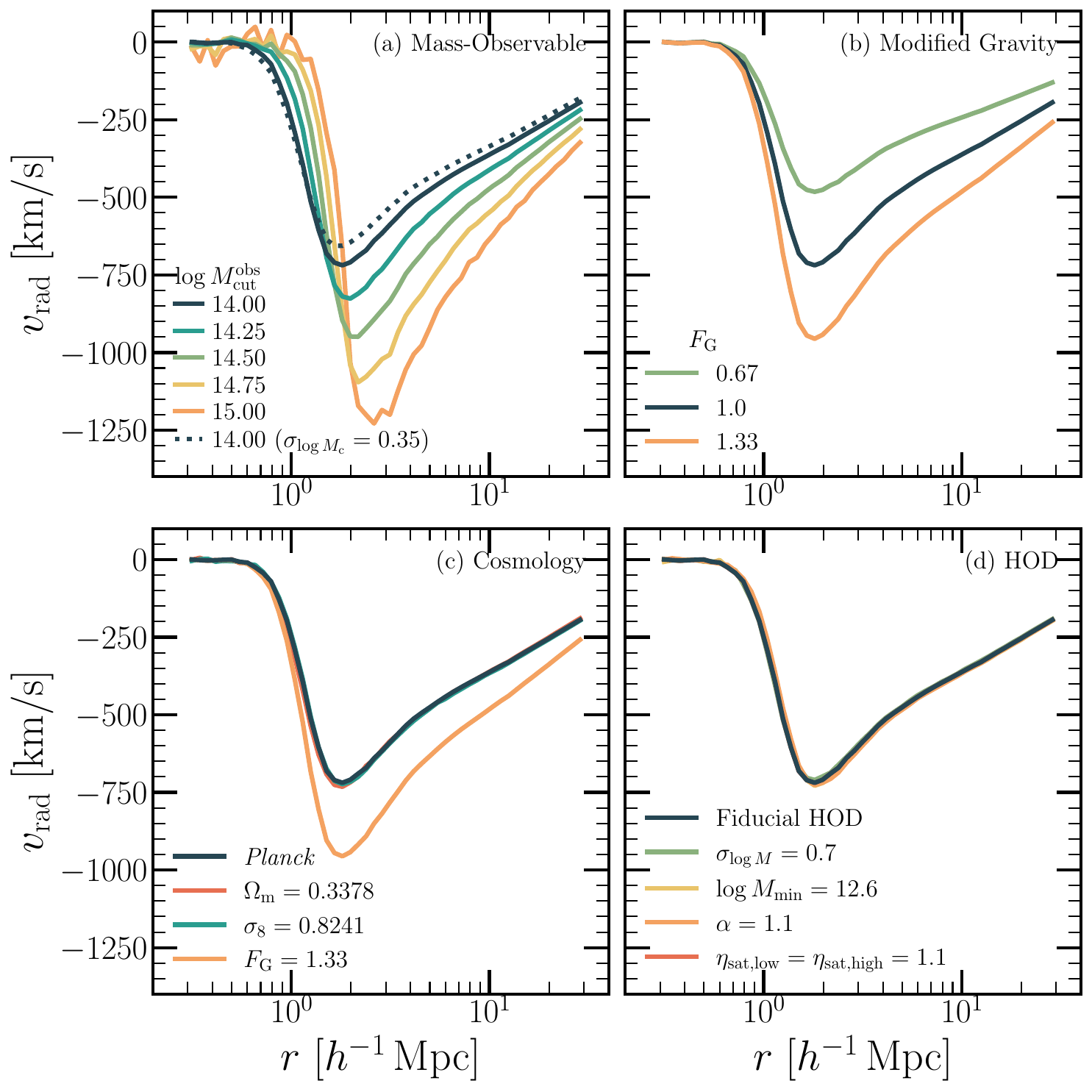}
\vspace{-5mm}
\caption{Mean radial velocity profile $\vrad(r)$ of galaxies around clusters in \AbacusSummit, averaged over all clusters in a single simulation phase (\texttt{ph000}) at the fiducial cosmology. The navy curve in each panel denotes the fiducial model. Panel~(a) shows variations in the mass--observable relation parameters $\lmcut$ and $\smscat$. Panel~(b) shows the response to the modified gravity parameter $\Fpar$. Panel~(c) shows variations in the cosmological parameters $\Om$ and $\sigma_8$. Panel~(d) shows variations in the HOD parameters. The profile responds most strongly to $\Fpar$ and the mass--observable parameters, and is comparatively insensitive to the range of HOD and cosmological parameter variations explored here.}
\label{fig:vrad}
\end{figure*}

\begin{figure}[t]
\centering
\includegraphics[width=\textwidth]
{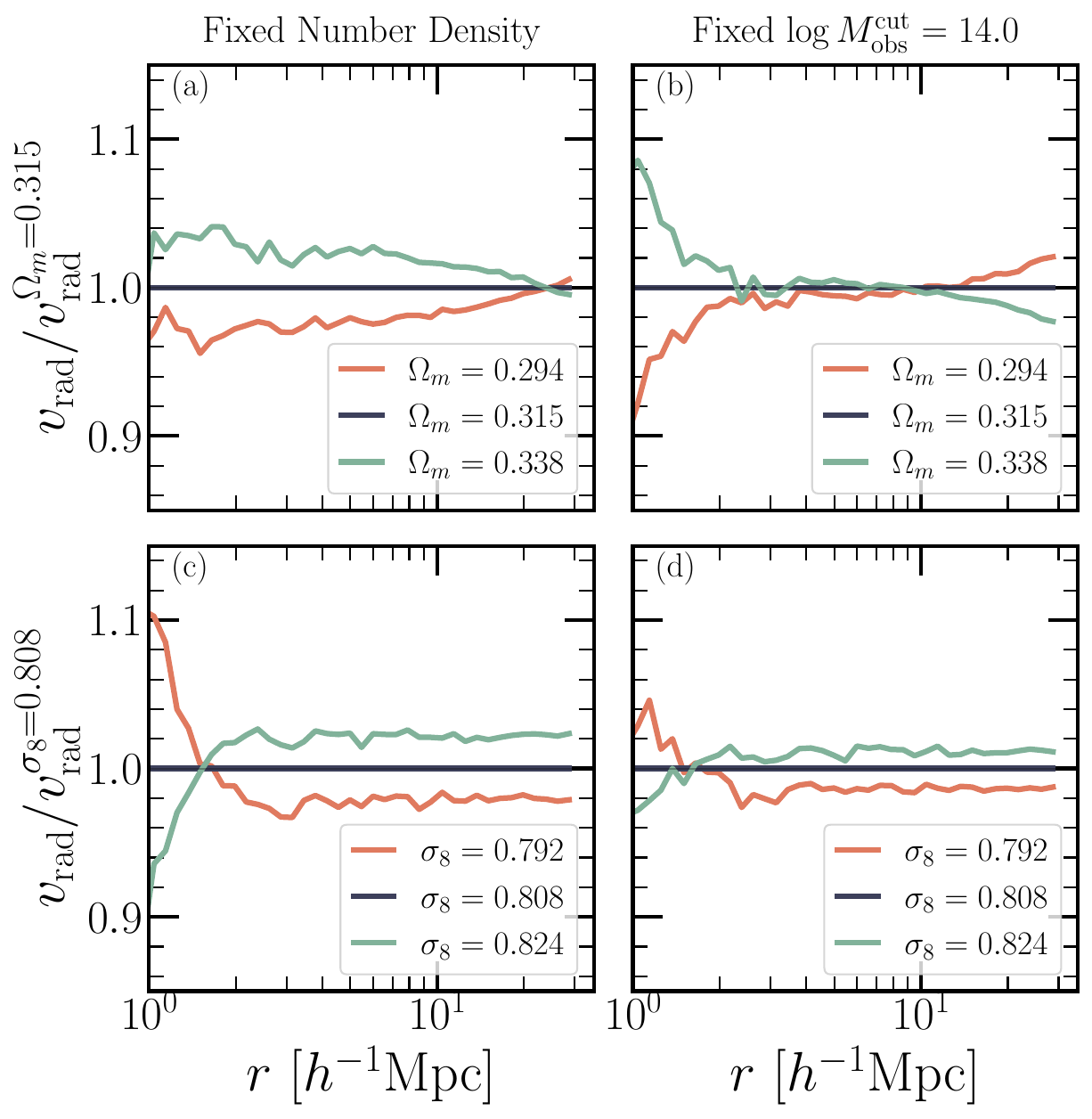}
\vspace{-5mm}
\caption{Ratio of the mean radial infall velocity profile to the fiducial, $\vrad / \vrad^{\rm fid}$, for variations in $\Om$ (top panels) and $\sig$ (bottom panels). The left column shows fixed-number-density samples (the $100{,}000$ most massive halos) and the right column shows samples at fixed $\lmcut = 14.0$. Orange and green curves indicate the low- and high-parameter variants, respectively, with the fiducial shown in navy. At fixed number density, deviations reach $\sim3\%$ at $r \approx 2$--$5\ h^{-1}\ \mathrm{Mpc}$, driven primarily by shifts in the selected halo mass. At fixed $\lmcut$, residual differences are $\lesssim 2\%$ at $r \gtrsim 2\ h^{-1}\ \mathrm{Mpc}$, demonstrating that $\vrad(r)$ outside the virialized region is largely insensitive to both $\Om$ and $\sig$ once the halo mass scale is held fixed. Larger deviations at $r \lesssim 2\ h^{-1}\ \mathrm{Mpc}$ reflect changes in the transition from infall to virialized motion.}
\label{fig:vrad_ratio}
\end{figure}

Figure~\ref{fig:vrad} presents $\vrad(r)$ computed from the halo and galaxy catalogs described in section~\ref{sec:sims}, evaluated at the fiducial \textit{Planck} cosmology using a single simulation phase (\texttt{ph000}). For the cosmological variants, we use a single matched phase from the corresponding boxes listed in table~\ref{tab:cosmo}. The profile exhibits the characteristic structure of the cluster infall regime. At large separations ($r \gtrsim 10\ h^{-1}\ \mathrm{Mpc}$), the mean radial velocity approaches zero, consistent with the background Hubble flow. Moving inward, galaxies accelerate toward the cluster, and the mean infall speed peaks at $r \sim 1$--$3\ r_{\rm vir}$. At smaller radii, coherent infall gives way to virialized, approximately isotropic motions within the cluster interior, and the mean radial velocity returns toward zero. Zu \& Weinberg~\cite{zuw2013} decompose cluster velocity fields into infalling and orbiting components, as have several subsequent studies \citep{Aung2021, Garcia2021, Wagoner2021, Garcia2023, Aung2023, Salazar2025, Shields2026, Sweeney2026}. 

Panel~(a) shows how $\vrad(r)$ responds to variations in the mass--observable relation parameters $\lmcut$ and $\smscat$. Increasing $\lmcut$ selects halos of higher virial mass, deeper gravitational potential, and larger virial radius, raising the peak infall speed and shifting the peak to larger radius. The peak infall speed increases from roughly $700\,\kms$ at the fiducial $\lmcut = 14.0$ (navy curve) to roughly $1250\,\kms$ at $\lmcut = 15.0$ (orange curve), reflecting the faster infall into more massive and spatially extended potential wells, consistent with the expected mass--velocity scaling. Increasing $\smscat$ at fixed $\lmcut$ (dotted navy curve, $\smscat = 0.35$~dex) introduces a larger fraction of lower-mass halos into the sample through upward scatter across the selection threshold, reducing the effective mean cluster mass and lowering the peak infall speed by roughly $50\,\kms$ relative to the fiducial.
 
Panel~(b) shows how $\vrad(r)$ responds to variations in $\Fpar$. Since $\Fpar$ rescales the mean halo velocities without altering the internal velocity dispersions, its effect on $\vrad(r)$ is nearly identical to a direct multiplication by $\Fpar$. An increase in $\Fpar$ from 1.0 to 1.33 raises the peak infall speed from approximately $700$ to $950\,\kms$, comparable to the effect of increasing $\lmcut$ from 14.0 to 14.5. However, the two effects are distinguishable: raising $\lmcut$ also shifts the peak to larger radius by increasing the halo virial radius, whereas increasing $\Fpar$ uniformly rescales the velocity field at fixed spatial structure, raising the peak infall speed without appreciably shifting its location.
 
Panel~(c) compares $\vrad(r)$ between the fiducial cosmology (navy curve) and the high-$\Om$ (red curve) and high-$\sig$ (blue curve) variants listed in table~\ref{tab:cosmo}. All three profiles are nearly indistinguishable, unlike the $\Fpar = 1.33$ curve (orange), whose peak infall speed is about $250\,\kms$ higher.

Figure~\ref{fig:vrad_ratio} shows the ratio $\vrad/\vrad^{\rm fid}$ for variations in $\Om$ (top row) and $\sig$ (bottom row). These ratios explain why $\vrad(r)$ is so insensitive to cosmology. For each parameter, we compare fixed-number-density samples (the $100{,}000$ most massive halos; left column) and samples selected at a fixed mass threshold, $\lmcut = 14.0$ (right column). This decomposition isolates changes in the velocity field due to the cosmological parameter itself from those arising indirectly through shifts in the selected cluster population. Figure~\ref{fig:vrad} corresponds to the fixed $\lmcut$ case.

When cluster abundance is held fixed (panel~(a)), a higher $\Om$ requires a larger mass threshold, since a denser universe produces more massive structures. The infall velocities around these more massive halos are enhanced by $2.9\%$ at $r=2\ h^{-1}\ \mathrm{Mpc}$, decreasing to $1.6\%$ by $r=10\ h^{-1}\ \mathrm{Mpc}$ (green curve, $\Om = 0.338$). At fixed $\lmcut$ (panel~(b)), the differences are small at $r \gtrsim 2\ h^{-1}\ \mathrm{Mpc}$ because, relative to the fiducial model, the high-$\Om$ model raises the mean mass of the selected halos by only $1.0\%$ ($0.0044$~dex) and the linear growth rate $f(\Om) \approx \Om^{0.55}$ at $z = 0.3$ by only $3\%$. Below $r \approx 2\ h^{-1}\ \mathrm{Mpc}$ the curves diverge more strongly, with the high-$\Om$ model up to $\sim 8\%$ above and the low-$\Om$ model up to $\sim 9\%$ below the fiducial near $r = 1\ h^{-1}\ \mathrm{Mpc}$. Lower $\Om$ also shifts matter-radiation equality to larger scales, boosting large-scale power and producing slightly larger $\vrad$ at $r \gtrsim 15\ h^{-1}\ \mathrm{Mpc}$, reaching a $2.1\%$ excess by $r=30\ h^{-1}\ \mathrm{Mpc}$ (the high-$\Om$ model is correspondingly $2.3\%$ lower).

Raising $\sig$ increases the amplitude of the matter power spectrum without changing its shape. At fixed number density (panel~(c)), a higher $\sig$ (green curve) selects more massive halos, producing enhanced infall velocities of $1.9$--$2.4\%$ at $r = 10$--$30\ h^{-1}\ \mathrm{Mpc}$, consistent with the $2.0\%$ fractional difference in $\sig$ itself. The ordering inverts at $r \lesssim 1.6\ h^{-1}\ \mathrm{Mpc}$, where the high-$\sig$ model falls below unity, by $\sim 9\%$ at $r = 1\ h^{-1}\ \mathrm{Mpc}$. This inversion arises because more massive halos have larger virial radii, shifting the transition from coherent infall to virialized orbital motion to larger physical scales. Fixing $\lmcut$ suppresses the difference between models to $\lesssim 1.4\%$ at $r \gtrsim 5\ h^{-1}\ \mathrm{Mpc}$ (panel~(d)), and the small-scale inversion persists at reduced amplitude. The remaining difference may be caused by the slight increase in the mean halo mass above the selection threshold, which is about $0.0078$~dex ($1.8\%$) higher in the high-$\sig$ model.
 
Returning to figure~\ref{fig:vrad}, panel~(d) shows the response of $\vrad(r)$ to variations in the HOD parameters $\siglogM$ (green), $\logMmin$ (yellow), $\alphaHOD$ (orange), and $\etasat$ (red), each evaluated at the perturbed values listed in table~\ref{tab:params}. None of these variations produces a discernible deviation from the fiducial profile, indicating that $\vrad(r)$ is insensitive to the details of the galaxy--halo connection captured by our HOD parameterization. This is broadly consistent with the expectation that gravitational acceleration at radius $r$ is primarily sourced by the cluster mass profile rather than by the internal structure of the galaxy population. Our HOD model does not include assembly bias, in which galaxy properties depend on halo formation history or large-scale environment beyond halo mass alone. Since such effects could preferentially select galaxies tracing different regions of the infall zone, we cannot entirely rule out that a more realistic treatment of assembly bias would introduce environmental sensitivity in $\vrad(r)$ that our current model does not capture. However, because $\vrad(r)$ is insensitive to the standard HOD parameters, which also change how galaxies trace the matter distribution, a significant impact of assembly bias on $\vrad(r)$ seems unlikely.
 
Taken together, the four panels show that $\vrad(r)$ responds most strongly to $\Fpar$, $\lmcut$, and $\smscat$, while remaining largely insensitive to the HOD and cosmological parameters explored here. Unlike $\lmcut$ and $\smscat$, which also shift the peak location, $\Fpar$ produces a spatially uniform velocity enhancement. HOD and cosmological parameters influence our observable primarily through the real-space correlation function and the number densities of galaxies and halos, rather than through $\vrad$ itself. 

\section{The Redshift-Space Galaxy--Cluster Cross-Correlation Function}\label{sec:xicgs}

\begin{figure}[t]
\centering
\includegraphics[width = 0.9\textwidth]{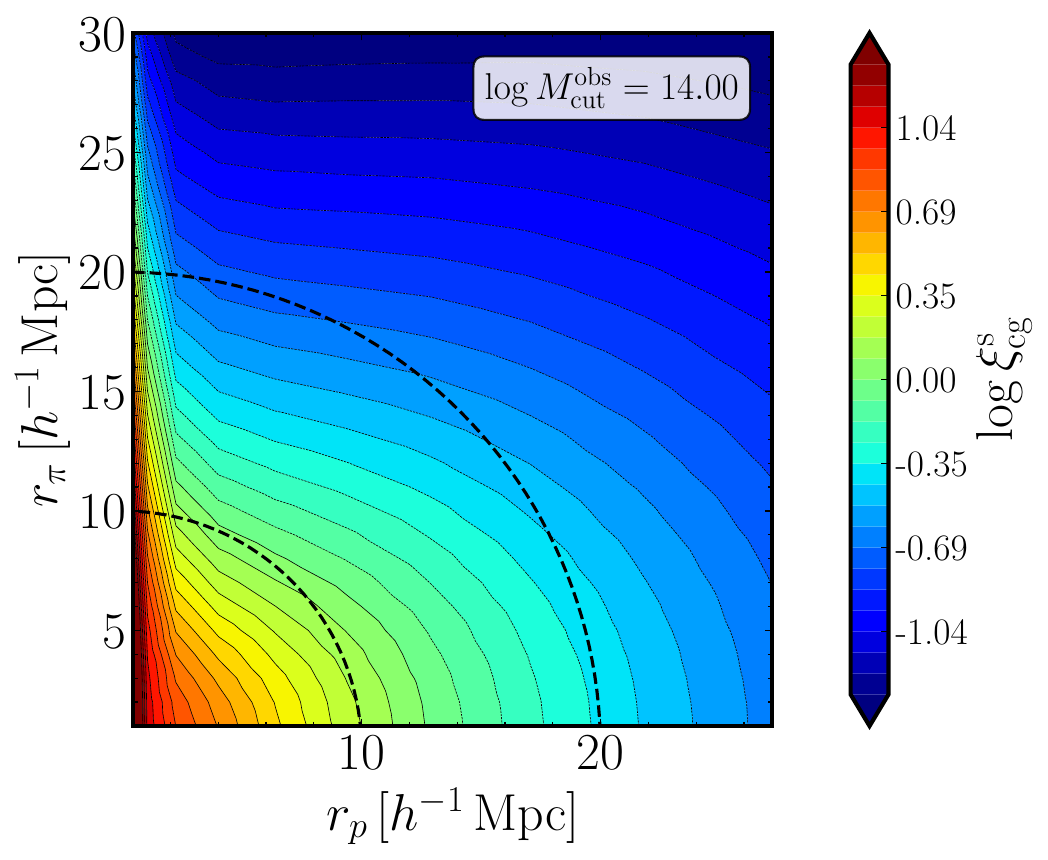}
\vspace{-5mm}
\caption{The redshift-space galaxy--cluster cross-correlation function, $\log_{10} \xicgsfull$, at $z = 0.3$ from \textsc{AbacusSummit}. This panel adopts fiducial HOD parameters with $\smscat = 0.25$ dex, $\Fpar = 1$, and $\log M_{\rm cut}^{\rm obs} = 14.00$. The vertical elongation at small $r_p$ reflects the FoG effect. Dashed circles of radius $r = 10$ and $20\,h^{-1}\,{\rm Mpc}$ provide an isotropic reference, highlighting the Kaiser \cite{Kaiser1987} compression along $r_\pi$ at larger scales.}
\label{fig:xicgs_rprpi}
\end{figure}

Having characterized how $\vrad(r)$ responds to variations in the model parameters, we now turn to the observable quantity that encodes this velocity field in spectroscopic data. The real-space infall profile is not directly accessible; instead, peculiar velocities displace the apparent line-of-sight positions of galaxies, imprinting a characteristic anisotropy on the redshift-space galaxy--cluster cross-correlation function $\xicgsfull$.

We estimate $\xicgsfull$ from galaxy--cluster pair counts as
\begin{equation}
\xicgsfull = \frac{DD(r_p,r_\pi)}{RR(r_p,r_\pi)} - 1,
\label{eq:estimator}
\end{equation}
where $DD(r_p,r_\pi)$ is the number of galaxy--cluster pairs in a given $(r_p,r_\pi)$ bin and $RR(r_p,r_\pi)$ is the expected number of pairs if both the galaxies and clusters were independently drawn from uniform random distributions with the same mean number densities. Assuming uniform densities, we compute the random counts analytically:
\begin{equation}
RR(r_p,r_\pi) = 2 \pi \, N_c \, n_g \, (r_{p,2}^2 - r_{p,1}^2) \, 
(r_{\pi,2} - r_{\pi,1}),
\label{eq:randoms}
\end{equation}
where $N_c$ is the number of clusters, $n_g$ is the galaxy number density, and the factor of 2 enters because values of $\pm \rp$ are assigned to the same bin. Since we compute the $RR$ term analytically and the simulations employ periodic boundary conditions, we do not require the Landy--Szalay estimator~\cite{Landy1993} to reduce variance.

We compute galaxy--cluster pair counts, $DD(\rp,\rpi)$, using the \texttt{Corrfunc} package \citep{Sinha2020, manodeep_sinha_2020_3634195} in cylindrical coordinates. The transverse separation spans $\rp \in [0.3, 30]\,h^{-1}\mathrm{Mpc}$ with a non-linear binning scheme,
\begin{equation}
\rp(u) = 0.3 + (30 - 0.3)\, u^{\alpha_{\rm bins}}, \quad u \in [0,1],
\end{equation}
where $\alpha_{\rm bins} = 2$ concentrates bins toward small scales where the signal varies most rapidly. We divide the line-of-sight separation into 30 uniform bins of width $1\,h^{-1}\mathrm{Mpc}$, covering $\rpi \in [0,30]\,h^{-1}\mathrm{Mpc}$.

We incorporate redshift-space distortions by displacing galaxy and halo positions along the line of sight,
\begin{equation}
r_\pi = r_{\parallel} + \frac{(1+z)}{H(z)}\,v_{\parallel},
\end{equation}
where $r_{\parallel}$ is the real-space line-of-sight separation, $v_{\parallel}$ is the peculiar velocity component along the line of sight, and $H(z)$ is the Hubble parameter. To increase the precision of the measurement, we treat each of the three Cartesian axes of the simulation box in turn as the line of sight and average the results, enforcing periodic boundary conditions by wrapping displaced positions back into the box. For $\Fpar \neq 1$, we multiply the halo peculiar velocities by $\Fpar$, and galaxies inherit these enhanced bulk motions. This rescales the mean infall profile (equation~\ref{eq:fpar}) without changing the internal velocity dispersions within halos.

Figure~\ref{fig:xicgs_rprpi} shows $\logxicgsfull$ at $z=0.3$, averaged over 25 independent realizations of the fiducial \AbacusSummit\ cosmology and over the three line-of-sight directions. In the absence of peculiar velocities, contours of constant $\xi_{cg}$ would be circular in the $(r_p, r_\pi)$ plane. Deviations from circular symmetry therefore directly reflect the influence of the velocity field. On small transverse scales, large virial velocities of galaxies within cluster halos elongate the contours along the line of sight, producing the FoG. On larger scales, coherent infall compresses the contours along $r_\pi$ \citep{Kaiser1987, Sargent1977}, a distortion commonly referred to as the Kaiser effect.

\section{Sensitivity of Observables to Model Parameters}
\label{sec:sensitivity}

To quantify how each model parameter affects our observables, we compute derivatives of the full data vector ($\logxicgs$, $\log n_{\rm gal}$, and $\log n_{\rm halo}$) with respect to each parameter $\theta$, evaluated at the fiducial values listed in table~\ref{tab:params}. We calculate these derivatives using the centered finite-difference approximation,
\begin{equation}
    \frac{\partial \log \mathcal{O}}{\partial \theta} \approx
    \frac{\log \mathcal{O}(\theta + \delta\theta) - 
    \log \mathcal{O}(\theta - \delta\theta)}{2\,\delta\theta},
    \label{eq:deriv}
\end{equation}
where $\mathcal{O}$ denotes any element of the data vector, and table~\ref{tab:params} lists the step sizes $\delta\theta$. For $\logxicgs$ and the number densities, we derive the HOD, modified gravity, and mass--observable parameter derivatives by averaging over the 25 \textit{Planck} fiducial cosmology phases of \AbacusSummit, and obtain the $\Omega_m$ and $\sigma_8$ derivatives from the paired simulation boxes listed in table~\ref{tab:cosmo}. We show the real-space counterparts of the $\logxicgs$ derivatives in appendix~\ref{app:realspace} for reference.

\subsection{The Redshift-Space Cross-Correlation Function}
\label{sec:deriv_xicgs}

\begin{figure*}[]
\centering
\includegraphics[width = \textwidth]{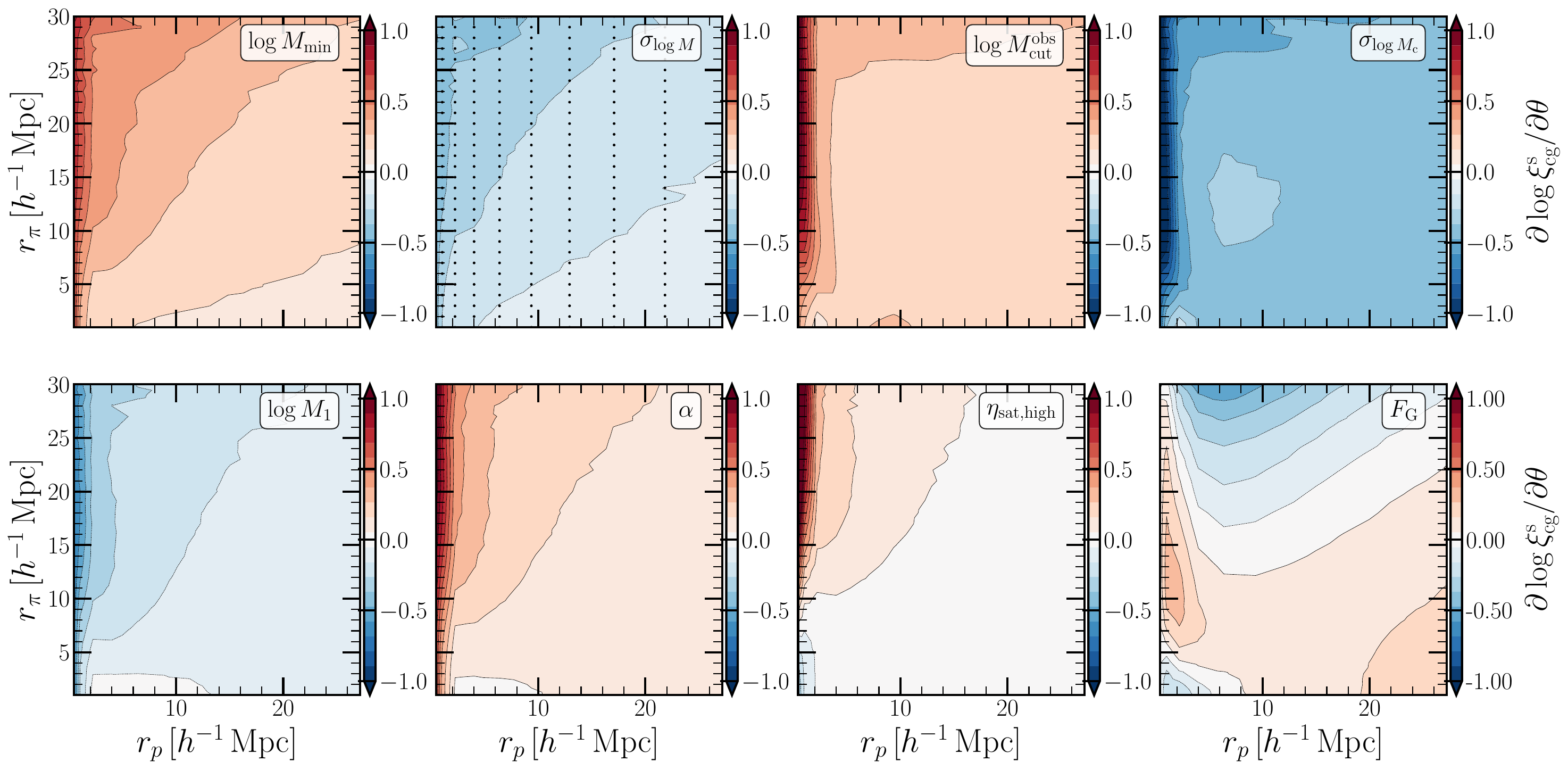}
\vspace{-5mm}
\caption{Derivatives $\partial \logxicgs / \partial \theta$ evaluated at the fiducial parameter values via centered finite difference for the HOD and mass selection parameters, measured at $z=0.3$. The parameter $\theta$ being varied is labeled in the upper right corner of each panel. Red hues denote a positive derivative, corresponding to an enhancement of the clustering signal, and blue hues denote a negative derivative, corresponding to a suppression of the signal. Each panel displays a distinct spatial pattern across the $(r_p, r_\pi)$ plane, reflecting the unique signature each parameter imprints on $\logxicgs$. Black dots in the $\siglogM$ panel indicate the underlying grid of $(r_p,r_\pi)$ bins at which the measurements are computed. The contours in all panels are interpolated between these points and are intended as a qualitative guide to the spatial structure of each derivative.}
\label{fig:2D_derivatives}
\end{figure*}

\begin{figure*}
\centering
\includegraphics[width = \textwidth]{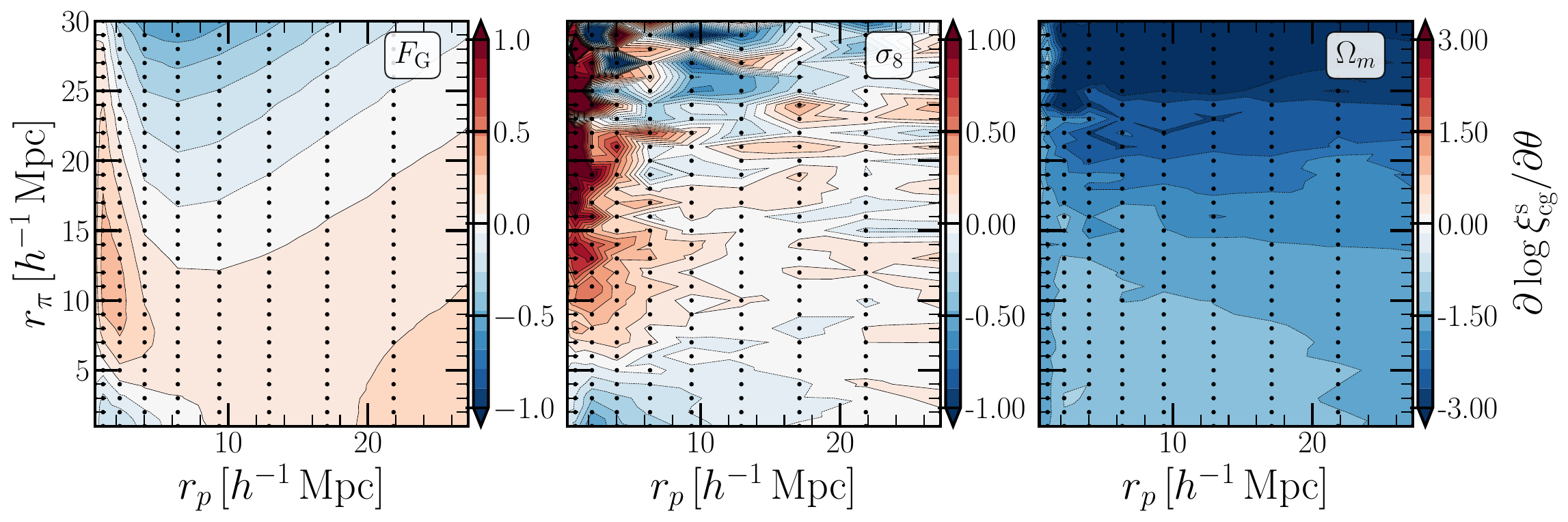}
\vspace{-5mm}
\caption{Derivatives $\partial \logxicgs / \partial \theta$, as in figure~\ref{fig:2D_derivatives}, but for the modified gravity parameter $\Fpar$ (left), clustering amplitude $\sigma_8$ (center), and matter density $\Omega_m$ (right), measured at $z=0.3$ using a halo sample selected at fixed $\lmcut$. The left and center panels share a common color scale; the right panel uses a separate scale due to the larger amplitude of the $\Omega_m$ derivative. The cosmological derivatives are visibly noisier as they are computed from single paired simulation phases rather than the full average over 25 realizations.}
\label{fig:cosmo_derivatives}
\end{figure*}

Figures~\ref{fig:2D_derivatives} and~\ref{fig:cosmo_derivatives} show $\partial \logxicgs / \partial \theta$ throughout the $(\rp, \rpi)$ plane. Figure~\ref{fig:2D_derivatives} presents derivatives with respect to the HOD and mass--observable parameters, while figure~\ref{fig:cosmo_derivatives} presents those with respect to the cosmological parameters, with $\Fpar$ included in both for reference. In each figure, red and blue hues indicate positive and negative derivatives, corresponding to enhancement and suppression of the clustering signal, respectively. The black dots in the $\siglogM$ panel of figure~\ref{fig:2D_derivatives} and in all panels of figure~\ref{fig:cosmo_derivatives} mark the discrete $(\rp, \rpi)$ bins at which we evaluate the derivatives. These values are the quantitative output used in the Fisher analysis of section~\ref{sec:fisher}. The contours interpolated between these points serve only to visualize the spatial structure of each derivative for the reader and play no role in the quantitative analysis itself. We present derivatives of the real-space correlation function $\xicgrfull$ in appendix~\ref{app:realspace} and reference them throughout the discussion below.

\subsubsection{Galaxy Occupation Parameters}
\label{sec:deriv_central}
The parameters $\logMmin$ and $\siglogM$ govern which halos host a central galaxy by setting the characteristic mass scale and the sharpness of the occupation transition (equation~\ref{eq:Ncen}). We interpret their effects on $\xicgs$ most directly through the way they shift the effective bias of the galaxy sample. On large scales, where linear bias applies, the galaxy--cluster cross-correlation approaches
\begin{equation}
    \xi_{\rm cg}(r) = b_g \, b_c \, \xi_m(r),
    \label{eq:xicg-linear}
\end{equation}
where $\xi_m(r)$ is the matter auto-correlation function, $b_c$ is the effective large-scale bias of the cluster sample, and $b_g$ is the effective large-scale bias 
of the galaxy sample, given by \citep{berlind2002}
\begin{equation}
    b_g = \bar{n}_g^{-1}
          \int_{M_{\rm min}}^{\infty}
          \frac{dn}{dM}\,\langle N (M)\rangle\,b_h(M)\,dM,
    \label{eq:bgal}
\end{equation}
where $dn/dM$ is the halo mass function, $\langle N (M)\rangle = \langle N_{\rm cen}(M)\rangle + \langle N_{\rm sat}(M)\rangle$ is the mean total occupation, $b_h(M)$ is the halo bias, and $\bar{n}_g$ normalizes the weighted average. Any HOD parameter that shifts the occupation-weighted mean of $b_h(M)$ directly alters $b_g$ and hence the large-scale amplitude of $\xicgs$.

Raising $\logMmin$ restricts central occupation to more massive, more strongly biased halos, increasing $b_g$ through equation~(\ref{eq:bgal}) and amplifying the two-halo clustering term through equation~(\ref{eq:xicg-linear}). Since these halos lie far above the $M_{\rm min}$ threshold, raising $\log M_{\rm min}$ does not affect the number of galaxies within clusters. However, it lowers the overall galaxy number density $n_{\rm g}$, which increases the one-halo term of $\xicgs$ by reducing the $RR$ denominator in equation~(\ref{eq:estimator}). This effect amplifies the derivative at small $r$ in real space (appendix~\ref{app:realspace}) and produces a strong positive response in the FoG regime (small $\rp$, wide range of $\rpi$) in redshift space (figure~\ref{fig:2D_derivatives}).

Increasing $\siglogM$ allows a larger fraction of lower mass, less biased halos to host central galaxies, producing effects nearly opposite to those of $\log M_{\rm min}$ in both real and redshift space. The primary difference is that the relative response in the one-halo FoG regime is weaker for $\siglogM$ than for $\log M_{\rm min}$. 

\subsubsection{Satellite Galaxy Occupation Parameters}
\label{sec:deriv_satellite}

The satellite occupation parameters $\logMone$ and $\alphaHOD$ govern the mean number of satellite galaxies per halo (equation~\ref{eq:Nsat}). Decreasing $\logMone$ increases the satellite count across halos of all masses, while increasing $\alphaHOD$ preferentially boosts satellites in high-mass halos relative to lower-mass halos. Both changes enhance the large-scale bias by increasing the contribution of high-mass, strongly biased halos in equation~(\ref{eq:bgal}), and they amplify the one-halo term by increasing the satellite population within clusters. The latter effect is stronger for $\alphaHOD$, since the clusters preferentially reside in high-mass halos where this parameter has the greatest leverage.

In both real space (appendix~\ref{app:realspace}) and redshift space (figure~\ref{fig:2D_derivatives}), the derivatives for $\logMone$ and $\alphaHOD$ resemble those for $\logMmin$, with a sign reversal for $\logMone$. While $\logMmin$ affects the one-halo term through the $RR$ denominator of equation~(\ref{eq:estimator}), $\logMone$ and $\alphaHOD$ act through the $DD$ numerator. The satellite parameters, particularly $\alphaHOD$, have a stronger influence on the one-halo term relative to the large-scale bias, producing a more pronounced FoG response. In particular, increasing $\alphaHOD$ produces a large enhancement of $\xicgsfull$ at small $\rp$ and large $\rpi$, since it increases the satellite population in the highest-mass clusters, which also have the largest velocity dispersions.

\subsubsection{Cluster Sample Selection Parameters}
\label{sec:deriv_cluster}
The parameters $\lmcut$ and $\smscat$ govern the properties of the cluster sample. Either increasing $\lmcut$ or decreasing $\smscat$ raises the mean cluster mass, so the two parameters have nearly opposite effects on $\xicgsfull$. More massive clusters have a larger bias factor $b_c$, and in real space the resulting amplification or suppression of the cross-correlation is nearly scale-independent beyond $r \approx 2\ h^{-1}\ \rm Mpc$. In the one-halo regime, the effect is stronger because more massive clusters host larger satellite populations, which in redshift space produce the FoG distortion. Consequently, the largest derivatives for the cluster parameters occur at small $\rp$ over a wide range of $\rpi$. Although more massive clusters also have larger infall velocities, the redshift-space derivatives at large $\rp$ show only marginally more structure than their real-space counterparts. The cluster selection parameters therefore affect $\xicgsfull$ primarily by modulating the FoG strength through the satellite population within clusters.

\subsubsection{Velocity Parameters}
\label{sec:deriv_velocity}
The velocity parameters modulate the line-of-sight motions of galaxies and halos without altering their positions, so derivatives of the real-space $\logxicgrfull$ vanish with respect to these parameters, and their signatures are entirely encoded in the anisotropy of the redshift-space $\logxicgsfull$. Figure~\ref{fig:2D_derivatives} presents derivatives with respect to $\etasathigh$ and $\Fpar$; we omit those with respect to $\etasatlow$ and $\etacen$, whose amplitudes are negligible compared to the other parameters. In the one-halo regime, $\etasatlow$ has no effect because it does not modulate cluster-mass halos, and $\etacen$ has little effect because it governs only the central galaxy. In the two-halo regime, these parameters affect $\xicgsfull$ in principle by modifying the line-of-sight velocity distribution (see equation~1 of ref.~\cite{zuw2013}), but since they do not alter the mean infall velocity, the impact of a slightly broader distribution proves negligible in practice.
 
The parameter $\etasathigh$ scales the velocity dispersions of satellite galaxies within cluster halos above $M = 10^{13.5}\,h^{-1}\,\Msun$ relative to the virial velocity dispersion. Increasing $\etasathigh$ broadens the line-of-sight velocity distribution of satellites orbiting within these massive halos, redistributing pairs along $\rpi$ within the one-halo regime. This produces a positive derivative at small $\rp$ and large $\rpi$, where pairs scatter to larger line-of-sight separations, and a negative derivative at small $\rp$ and small $\rpi$, where those same pairs are depleted, resulting in the dipole structure visible at small $\rp$ in figure~\ref{fig:2D_derivatives}. The derivative remains non-zero at large $\rp$ and large $\rpi$ because a significant fraction of excess pairs in this regime involves the high-velocity tail of the cluster population, but it vanishes at large $\rp$ and smaller $\rpi$. We emphasize that $10$--$20\%$ departures from $\etasathigh=1$ are entirely plausible within GR as a consequence of orbital anisotropy, dynamical friction, preferential disruption of satellites on low eccentricity orbits, or departures from dynamical equilibrium.
 
The modified gravity parameter $\Fpar$ enhances the coherent infall velocity field at fixed cluster mass and internal velocity dispersion, mimicking a gravitational fifth force that accelerates matter toward cluster overdensities without disturbing the cluster's virial equilibrium. Its signature is therefore confined to large-scale galaxy streaming, with no corresponding change in the FoG signal. A residual response does appear at small $\rp$, but this arises from galaxy pairs in the two-halo regime rather than from pairs within the cluster, as is the case for $\etasathigh$. Increasing $\Fpar$ shifts a fraction of these small-$\rp$ pairs from small to larger $\rpi$. At large $\rp$, compression of structure along the line of sight produces a positive derivative at small to intermediate $\rpi$ across all transverse scales. In contrast, at large $\rpi$, where pairs are drawn inward from large line-of-sight separations, the derivative becomes negative. Appendix \ref{app:model} presents an idealized spherical model for cluster redshift-space distortions, which explains the qualitative features of the $\Fpar$ derivative in figure~\ref{fig:2D_derivatives} but does not reproduce it quantitatively. 

\subsubsection{Cosmological Parameters}
\label{sec:deriv_cosmology}
Figure~\ref{fig:cosmo_derivatives} shows the derivatives with respect to $\sigma_8$ and $\Omega_{\rm m}$, alongside the $\Fpar$ derivative included for reference. We present the corresponding real-space derivatives in appendix~\ref{app:realspace}. As noted above, we compute the cosmological derivatives from single paired simulation phases rather than the full average over 25 realizations. They are consequently much noisier than the HOD and mass--observable derivatives in figure~\ref{fig:2D_derivatives}. Since the volume of a single \textsc{AbacusSummit} box is approximately ten times larger than the DESI BGS volume in the range $z = 0.25$--$0.35$, we expect this noise to have a small effect relative to the forecast statistical uncertainties. Nonetheless, a more robust assessment will require analysis of additional simulations. We consider our $\Omega_{\rm m}$ and $\sigma_8$ forecasts preliminary and subject to revision in future work.
 
The derivative with respect to $\sigma_8$ exhibits an anisotropic structure concentrated primarily in the one-halo regime. At small $\rp$, the topology resembles the dipole-like signature of $\etasathigh$, with a negative derivative at small $\rpi$ and a positive derivative at larger $\rpi$. This structure reflects the response of the cluster population to increasing $\sigma_8$ at fixed $\log M_{\rm cut}$. While a higher $\sigma_8$ globally enhances matter clustering, it also shifts the halo mass function toward more massive clusters above the selection threshold. These more massive clusters have deeper potential wells and larger internal velocity dispersions, driving stronger FoG stretching along $\rpi$ and redistributing pairs from small to large $\rpi$, locally competing with and exceeding the global amplitude boost. On large scales, the real- and redshift-space derivatives lack coherent structure and likely reflect noise from the limited simulation volume.
 
The derivative with respect to $\Omega_m$ is negative across both the one-halo and two-halo regimes. Because the cosmological parameters are defined at $z=0$, increasing $\Omega_m$ at fixed $\sigma_8(z=0)$ results in a lower growth factor at the epoch of observation ($z=0.3$). This suppresses the amplitude of matter fluctuations, reducing $\logxicgs$ throughout all scales and producing the negative derivative visible in figure~\ref{fig:cosmo_derivatives}. The derivatives are qualitatively similar in real and redshift space, suggesting that changes in 3D clustering, rather than redshift-space distortions, primarily drive them. This inference is also supported by figure~\ref{fig:vrad}c, which shows that $\vrad(r)$ is only weakly sensitive to changes in $\Om$ and $\sig$. However, the shapes of derivative contours for $\Om$ are somewhat different in real and redshift space. 
 
An important feature of figure~\ref{fig:cosmo_derivatives} is that the $\sigma_8$ and $\Omega_{\rm m}$ derivatives exhibit a different structure from that of $\Fpar$. Furthermore, $\sigma_8$ and $\Omega_{\rm m}$ influence the cluster space density (section~\ref{sec:deriv_nd}), while $\Fpar$ does not. Together, these distinctions suggest that little degeneracy exists between $\Fpar$ and the cosmological parameters $\sigma_8$ and $\Omega_{\rm m}$.

\subsection{Galaxy and Cluster Number Densities}
\label{sec:deriv_nd}

\begin{figure}[]
\centering
\includegraphics[width=\columnwidth]{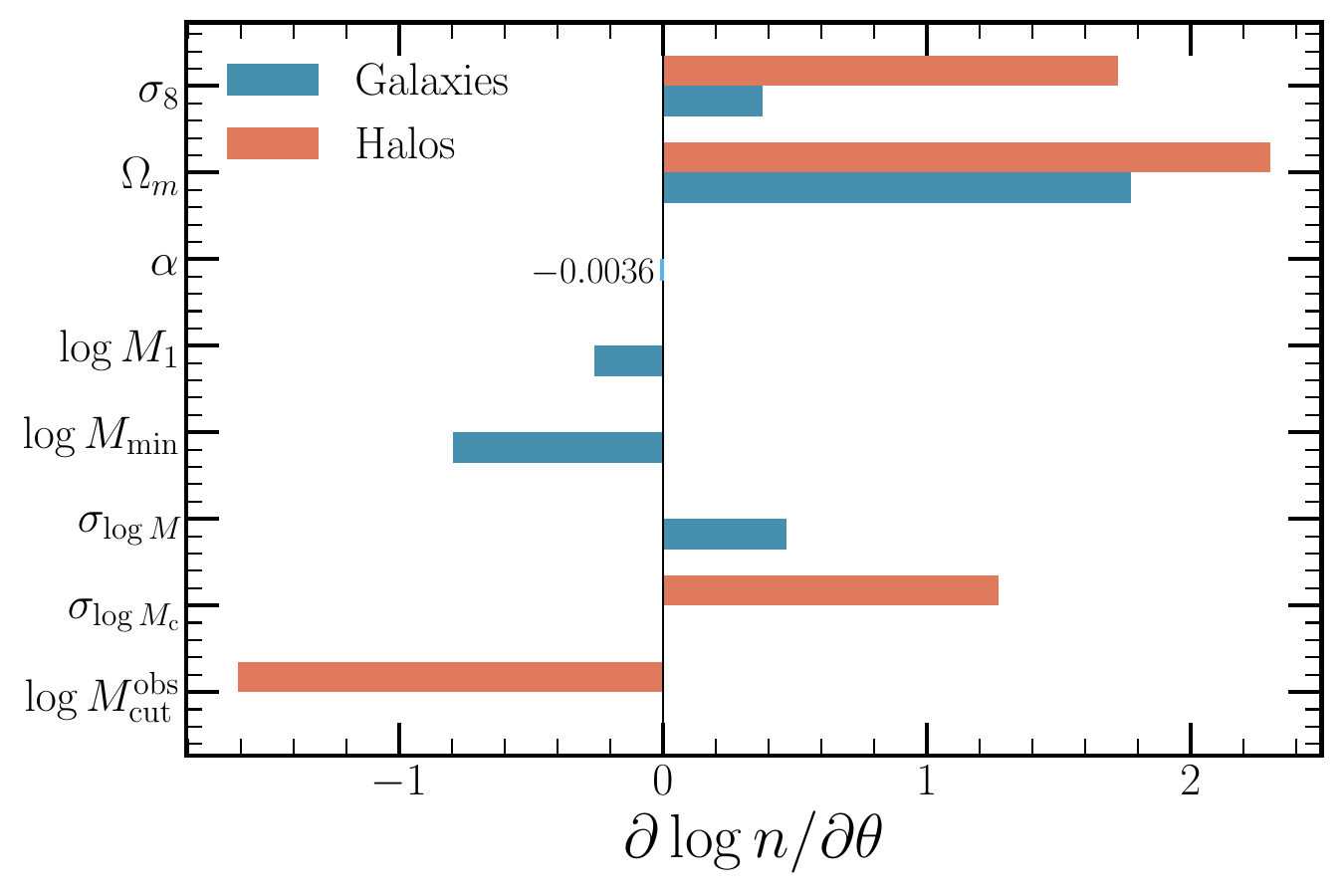}
\vspace{-3mm}
\caption{Derivatives $\partial \log n / \partial \theta$ of the log galaxy number density (blue) and log halo number density (red) with respect to each model parameter, evaluated at the fiducial parameter values. The velocity bias parameters ($\etasathigh$, $\etasatlow$, $\etacen$) and the modified gravity parameter $\Fpar$ are omitted as they have no impact on halo or galaxy number densities.}
\label{fig:nd_derivatives}
\end{figure}

In addition to $\logxicgs$, we include $\log n_{\rm gal}$ and $\log n_{\rm halo}$ in our data vector, calculated directly from the simulation as $n_{\rm gal} = N_{\rm gal}/V_{\rm box}$ and $n_{\rm halo} = N_{\rm halo}/V_{\rm box}$. Figure~\ref{fig:nd_derivatives} shows their derivatives with respect to each model parameter. We average the derivatives with respect to the HOD and mass--observable parameters over the 25 \textit{Planck} fiducial cosmology phases, and compute those with respect to $\Omega_m$ and $\sigma_8$ using the paired simulation boxes in table~\ref{tab:cosmo}. We omit $\Fpar$ and the velocity bias parameters ($\etasathigh$, $\etasatlow$, $\etacen$) from figure~\ref{fig:nd_derivatives}, since they modify the motions of galaxies and halos without affecting their total counts.

The halo number density ($\log n_{\rm halo}$; shown in red) responds only to the cluster selection and cosmological parameters, since HOD variations only change the distribution of galaxies within halos and have no effect on the halo population itself. Increasing $\lmcut$ restricts the cluster sample to rarer and more massive halos, producing a large negative derivative. Increasing $\smscat$ allows more low-mass halos to scatter above the selection threshold than high-mass halos to scatter below it, resulting in a net increase in $n_{\rm halo}$. Both $\sigma_8$ and $\Omega_m$ produce positive derivatives, as higher values of either parameter increase the overall abundance of halos above the selection threshold (see refs.~\cite{White1993, Zheng2002} for a more detailed discussion). For $\sigma_8$, the enhanced clustering amplitude shifts the halo mass function toward more massive halos, directly increasing the abundance above $\lmcut$. For $\Omega_m$, the increased background matter density shifts the mass function to higher mass at all space densities, which outweighs the slight suppression of the growth factor at $z = 0.3$.

The galaxy number density ($\log n_{\rm gal}$; shown in blue) is sensitive to both the HOD and cosmological parameters, since both govern the abundance of halos available to host galaxies. It is insensitive to $\lmcut$ and $\smscat$, as these parameters determine which halos enter the cluster sample but do not affect galaxy assignment. Among the HOD parameters, $\logMmin$ and $\siglogM$ produce the dominant responses, reflecting the fact that central galaxies constitute approximately 80\% of the total galaxy population. Raising $\logMmin$ restricts central occupation to more massive halos, leaving a large population of lower-mass halos without central galaxies and producing a large negative derivative. Increasing $\siglogM$ broadens the occupation transition, allowing more lower-mass halos to host central galaxies and yielding a positive derivative of comparable magnitude. The satellite parameters $\logMone$ and $\alphaHOD$ have a smaller impact on $\log n_{\rm gal}$, as satellites constitute only $\sim20\%$ of the total galaxy population. Raising $\logMone$ suppresses satellite occupation in intermediate mass halos, reducing $\log n_{\rm gal}$ and producing a negative derivative. Increasing $\alphaHOD$ steepens the satellite occupation--mass relation, redistributing satellites from abundant intermediate-mass halos to rare high-mass halos. The gain in the high-mass tail does not compensate for the loss at lower masses, yielding a small negative derivative. As with $\log n_{\rm halo}$, both $\sigma_8$ and $\Omega_m$ yield positive derivatives, driven by the increased abundance of halos massive enough to host central and satellite galaxies.

The derivatives in figures~\ref{fig:2D_derivatives}, \ref{fig:cosmo_derivatives}, and~\ref{fig:nd_derivatives} demonstrate that each parameter imprints a distinct signature across the full data vector. In particular, while the cosmological and HOD parameters have minimal impact on $\vrad(r)$ as shown in figure~\ref{fig:vrad}, their derivatives with respect to $\logxicgs$ exhibit structured anisotropic patterns across the $(\rp, \rpi)$ plane, indicating that the redshift-space clustering signal encodes additional parameter sensitivity beyond the mean infall velocity profile.

Of particular relevance for constraining $\Fpar$ is the degree to which other parameters produce similar signatures in $\logxicgs$. Parameters such as $\logMmin$, $\lmcut$, and $\alpha$ all produce broadly positive derivatives throughout the $(\rp,\rpi)$ plane, superficially similar to $\Fpar$. However, these parameters also enhance the FoG elongation at small $\rp$ through their impact on the halo mass and satellite population, whereas $\Fpar$ leaves the internal velocity dispersions unchanged and produces no corresponding FoG enhancement. Furthermore, $\Fpar$ produces a negative derivative at large $\rpi$ that is absent in the HOD and mass--observable derivatives. These distinctions in the anisotropic structure of $\logxicgs$ allow us to constrain $\Fpar$ independently of the mass--observable, HOD, and cosmological parameters, as we quantify in the following section.

\section{Fisher Forecast}\label{sec:fisher}
\subsection{Formalism and Construction}
\label{sec:fisher_construction}
We define the full data vector as
\begin{equation}
    \mathbf{d} =
    \left\{
    \log\xi_{\rm cg}^s(r_{p}, r_{\pi}),
    \ \log n_{\rm gal},
    \ \log n_{\rm halo}
    \right\},
\end{equation}
where the first set of elements corresponds to the two-dimensional correlation function evaluated over all $(r_p, r_\pi)$ bins, followed by the scalar observables $\log n_{\rm gal}$ and $\log n_{\rm halo}$. As described in section~\ref{sec:sensitivity}, we calculate the derivatives of each observable with respect to the model parameters using the centered finite-difference approximation in equation~(\ref{eq:deriv}).

Given the derivatives $\partial \mathbf{d} / \partial\theta_i$ evaluated at the fiducial parameter values (see tables~\ref{tab:cosmo} and \ref{tab:params}) and the covariance matrix $C$ of the full data vector (see section~\ref{sec:covariance} below), we construct the Fisher information matrix following the standard formalism 
\citep{tegmark1997},
\begin{equation}
    F_{ij} = \sum_{k,l}
              \frac{\partial d_k}{\partial \theta_i}\,
              \bigl(C^{-1}\bigr)_{kl}\,
              \frac{\partial d_l}{\partial \theta_j},
    \label{eq:fisher}
\end{equation}
where $d_k$ runs over all elements of the combined data vector. The covariance matrix takes the block form
\begin{equation}
    C =
    \begin{pmatrix}
        C_{\xi\xi} & C_{\xi n_{\rm g}} & C_{\xi n_{\rm h}} \\
        C_{n_{\rm g}\xi} & \sigma_{n_{\rm g}}^2 & C_{n_{\rm g} n_{\rm h}} \\
        C_{n_{\rm h}\xi} & C_{n_{\rm h} n_{\rm g}} & \sigma_{n_{\rm h}}^2
    \end{pmatrix},
    \label{eq:full_cov}
\end{equation}
where $C_{\xi\xi}$ is the covariance matrix of $\log\xi_{\rm cg}^s$, $\sigma_{n_{\rm g}}^2$ and $\sigma_{n_{\rm h}}^2$ are the variances of $\log n_{\rm gal}$ and $\log n_{\rm halo}$, the off-diagonal blocks $C_{\xi n_{\rm g}}$ and $C_{\xi n_{\rm h}}$ describe the cross-covariance between the correlation function and each number density, and $C_{n_{\rm g} n_{\rm h}}$ captures the covariance between the two number density observables. We obtain marginalized $1\sigma$ constraints on each parameter $\theta_i$ from the inverse Fisher matrix,
\begin{equation}
    \sigma(\theta_i) = \sqrt{\bigl(F^{-1}\bigr)_{ii}}.
    \label{eq:marginal_sigma}
\end{equation}
Unless otherwise stated, all quoted constraints are marginalized over the full parameter set.

\subsection{Covariance matrix}
\label{sec:covariance}

We estimate $C_{\xi\xi}$ from subvolumes of the 25 \textit{Planck} fiducial cosmology phases of \AbacusSummit. For each phase, we divide the simulation volume into $64$ equal subvolumes, each with side length $500\,h^{-1}\,\mathrm{Mpc}$, and measure $\logxicgsfull$ along the three Cartesian axes of the simulation box. We confirm that this choice of $25$ phases and $64$ subvolumes per phase, for a total of $N_s = 1600$ subvolumes, is sufficient by testing the convergence of the derived parameter constraints as a function of sample size, finding that all marginalized parameter errors at $N_s \approx 1000$ lie within $\sim 3\%$ of their $N_s = 1600$ values, and within $1.3\%$ at $N_s = 1280$. We compute a covariance matrix for each projection separately using the scatter across all subvolumes and phases,
\begin{equation}
    C_{ij} = \left\langle \left( d_i - \bar{d}_i \right)\left( d_j - \bar{d}_j \right) \right\rangle,
    \label{eq:cov_def}
\end{equation}
where $d_i$ is $\logxicgsfull$ in bin $i$ for a given subvolume and $\bar{d}_i$ is its mean over all subvolumes and phases. The combined covariance matrix is the average of the three projection-specific covariances,
\begin{equation}
    C = \frac{1}{3}\left(C_x + C_y + C_z\right).
    \label{eq:cov_avg}
\end{equation}

The \DESI\ BGS sample covers approximately $14{,}000\,\mathrm{deg}^2$ over $0.25 < z < 0.35$. In a \textit{Planck} cosmology, this corresponds to an effective cubic volume with side length $L_{\rm survey} = 912.55\,h^{-1}\,\mathrm{Mpc}$. Because each subvolume is smaller than the survey volume, we rescale the covariance by the ratio of subvolume to survey volume.

The values of $\sigma_{n_{\rm g}}$ and $\sigma_{n_{\rm h}}$ computed from the subvolumes are $0.0030$ and $0.0065$~dex, respectively, after scaling to the DESI BGS volume. Although the cosmic variance scatter in $\log n_{\rm gal}$ and $\log n_{\rm halo}$ is very small, the number densities in realistic surveys have systematic uncertainties from selection effects (e.g., photometric variation or spectroscopic incompleteness) and even from imperfect definition of the survey footprint (e.g., area masked by bright galaxies). We elect to be conservative and set both to $5\%$, adopting $\sigma_{n_{\rm g}} = \sigma_{n_{\rm h}} = 0.022$~dex in our fiducial forecast. We demonstrate in section~\ref{sec:variations} that our forecasts for $F_{\rm G}$ and the cosmological parameters are insensitive to the values of $\sigma_{n_{\rm g}}$ and $\sigma_{n_{\rm h}}$. Varying both from $1\%$ to $10\%$ changes the error on $F_{\rm G}$ by only $3\%$ and those on $\Om$ and $\sig$ by $\lesssim 1.5\%$ (see figure~\ref{fig:forecast_summary}). The constraints on $\log M_{\rm cut}$, $\sigma_{\log M}$, and $\log M_{\rm min}$, which are set primarily by the number densities, scale more directly with the assumed uncertainty. Because we are implicitly assuming that $\sigma_{n_{\rm g}}$ and $\sigma_{n_{\rm h}}$ are inflated by factors not represented in our simulations, we set the cross-covariance terms $C_{\xi n_{\rm g}}$ and $C_{\xi n_{\rm h}}$ to zero. We verify that this choice changes the forecast errors on $F_{\rm G}$, $\Om$, and $\sig$ by $\lesssim 1\%$ and on most HOD parameters by $\lesssim 2\%$; the largest effects are on $\sigma_{\log M}$ and $\log M_{\rm min}$, whose errors increase by $6\%$ and $9\%$, so neglecting these terms is conservative. We retain $C_{n_{\rm g} n_{\rm h}}$, the covariance between the two number densities, rescaled from its measured subvolume value to preserve their measured correlation coefficient ($\rho \approx 0.87$) under the assumed $5\%$ variances. Neglecting $C_{n_{\rm g} n_{\rm h}}$ would change the errors on $F_{\rm G}$, $\Om$, and $\sig$ by less than $1\%$ but shift those on $\log M_{\rm min}$ and $\log M_1$ by $\sim 7\%$ in opposite directions. In contrast, the off-diagonal structure of $C_{\xi\xi}$ is essential. A diagonal covariance would underestimate the error on $F_{\rm G}$ by $24\%$, shrink that on $\Om$ by a factor of $2.6$, and inflate that on $\sig$ by $42\%$.

With $C_{\xi n_{\rm g}}$ and $C_{\xi n_{\rm h}}$ set to zero, the covariance matrix in equation~(\ref{eq:full_cov}) becomes block diagonal, with $C_{\xi\xi}$ in one block and the two number density variances and their covariance in the other. We apply the Hartlap correction \cite{hartlap2007} to debias the inverse of the $C_{\xi\xi}$ block only,
\begin{equation}
    C^{-1}_{\xi\xi,\,\rm unbiased} = \frac{N_s - N_d - 2}{N_s - 1}\, C^{-1}_{\xi\xi},
    \label{eq:hartlap}
\end{equation}
where $N_s = 1600$ ($25$ phases $\times$ $64$ subvolumes) and $N_d = 300$ ($10$ bins in $r_p$ $\times$ $30$ bins in $r_\pi$), giving a correction factor of $1298/1599 \approx 0.812$. We do not apply the Hartlap correction to the number density block, since its values are set to an assumed external uncertainty rather than estimated from a finite number of realizations. We invert the two blocks independently and combine them to form the full inverse covariance used in equation~(\ref{eq:fisher}).

In a small number of bins at large $r_\pi$, $\xi^{\rm s}_{\rm cg} \le 0$ in some subvolumes. We set $\log\xi$ to a floor value in these cases, which inflates the variance of these bins and effectively removes them from the Fisher analysis.

\begin{figure}[tbp]
\centering
\includegraphics[width=\textwidth]{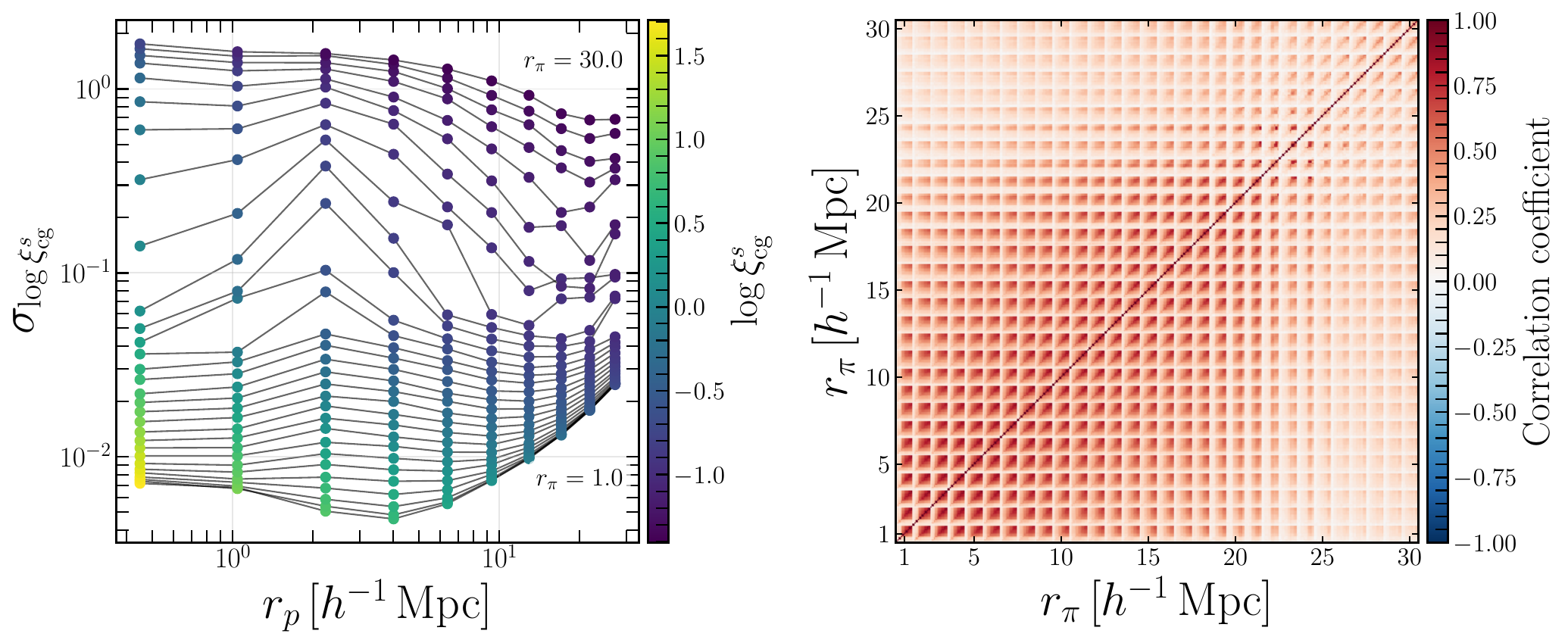}
\caption{Structure of the covariance and correlation matrices for $\log \xi_{\rm cg}^{\rm s}$, measured across the 1600 simulation subvolumes and rescaled to the DESI BGS volume. \textbf{Left:} the standard deviation $\sigma_{\log \xi_{\rm cg}^{\rm s}}$ as a function of transverse separation $r_p$. Each connected line corresponds to a fixed $r_\pi$ slice, ranging from $r_\pi = 1\,h^{-1}\,\mathrm{Mpc}$ (bottom) to $r_\pi = 30\,h^{-1}\,\mathrm{Mpc}$ (top), with points colored by the corresponding value of $\log \xi_{\rm cg}^{\rm s}$. The largest values ($\sigma \gtrsim 0.1$) occur in bins where $\xi^{\rm s}_{\rm cg} \le 0$ in some subvolumes (see text). \textbf{Right:} the full $300 \times 300$ correlation matrix of $\log \xi_{\rm cg}^{\rm s}$, ordered into 30 sequential $r_\pi$ slices, with the 10 $r_p$ bins increasing from left to right within each slice.}
\label{fig:covar}
\end{figure}

Figure~\ref{fig:covar} shows the covariance structure of $\logxicgs$. The left panel shows $\sigma_{\log\xi_{\rm cg}^{\rm s}}$ as a function of $r_p$, where each curve corresponds to a fixed $r_\pi$ slice and is colored by the value of $\logxicgs$ in that bin. At fixed $r_p$, $\sigma_{\log\xi_{\rm cg}^{\rm s}}$ increases with $r_\pi$, reflecting the decrease in the signal with line-of-sight separation. At small $r_\pi$, where the signal is strong, the uncertainty is below $0.03$~dex and rises only gradually towards large $r_p$. Once $\rpi$ exceeds $\sim 20\,h^{-1}\,\mathrm{Mpc}$, the error grows rapidly in all $r_p$, most significantly in the smallest $r_p$, where $\xi^{\rm s}_{\rm cg}$ beyond the extent of the FoG is close to zero and becomes negative in some subvolumes. The derivatives in figure~\ref{fig:2D_derivatives} exhibit distinctive behavior in this regime, but these bins carry little weight in the constraints because of their large statistical errors.

The right panel of figure~\ref{fig:covar} shows the correlation matrix of $\logxicgs$. The repeating block structure reflects the organization of the data vector into $r_\pi$ slices. Bins at the same $r_p$ are strongly correlated across different $r_\pi$ slices out to $r_\pi \sim 20\,h^{-1}\,\mathrm{Mpc}$, since they sample the same halos and their surrounding structure. These correlations weaken at larger $r_\pi$, where the bins are dominated by noise.

\subsection{Parameter constraints}
\label{sec:results}

\begin{figure}[tbp]
\centering
\includegraphics[width=\textwidth]{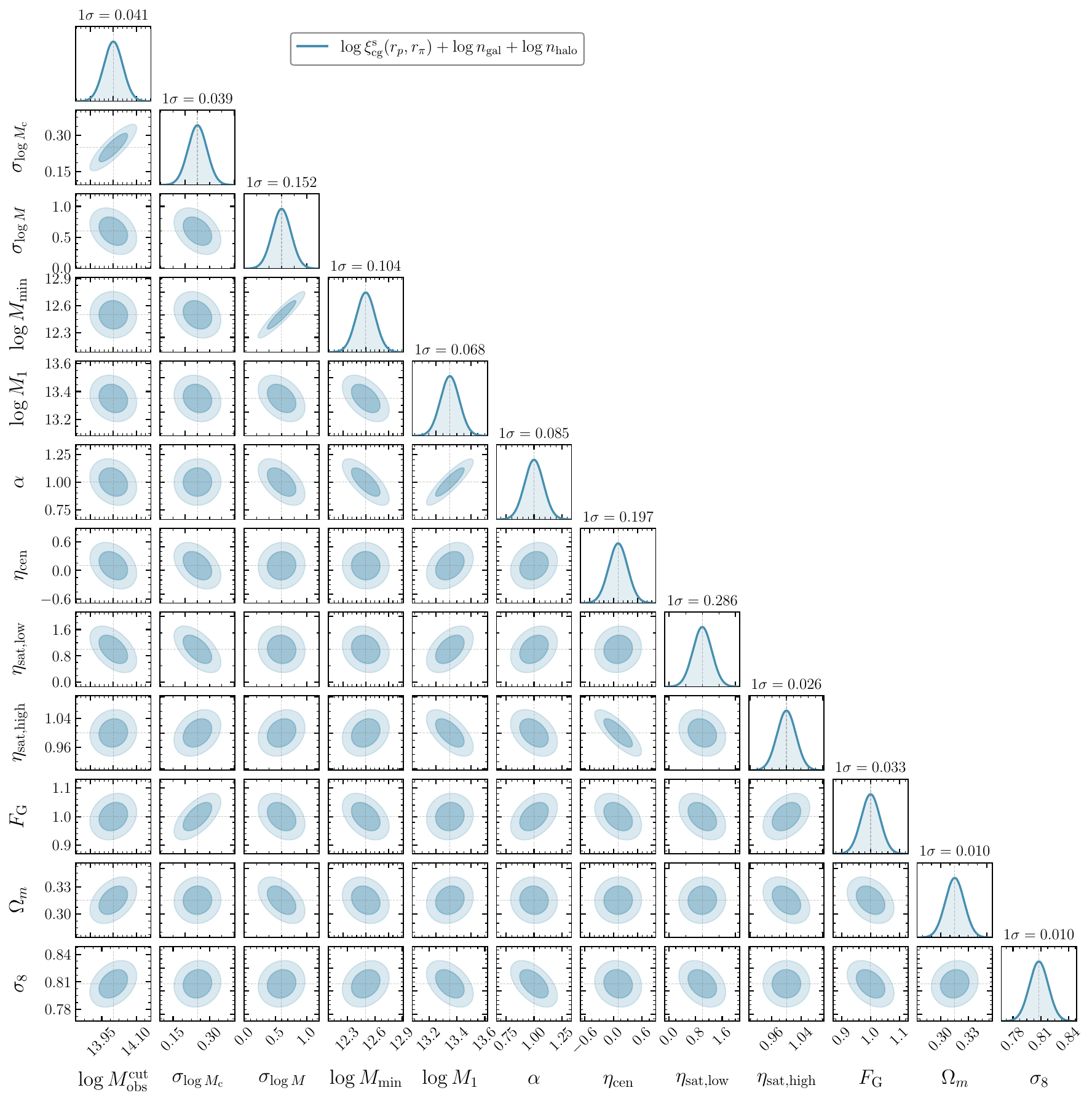}
\caption{Marginalized $1\sigma$ and $2\sigma$ posterior ellipses for the Full Forecast configuration, combining $\logxicgsfull$ with $\log n_{\rm gal}$ and $\log n_{\rm halo}$. The shaded ellipses correspond to the 68.3\% and 95.4\% confidence regions of the joint two-parameter distributions. The diagonal panels show the marginalized one-dimensional posterior for each parameter, with the fiducial value indicated by the dashed vertical line and the $1\sigma$ marginalized constraint quoted above each panel. From left to right, columns show the two cluster selection parameters, the four HOD occupation parameters, the three HOD parameters describing the velocity bias of central and satellite galaxies, the modified gravity parameter $\Fpar$, and the cosmological parameters $\Om$ and $\sig$. The modified gravity parameter $\Fpar$ is constrained to $\sigma(\Fpar) = 0.033$ after marginalization over the HOD parameters, demonstrating that the infall velocity signal can be isolated from galaxy--halo uncertainties.}
\label{fig:full_corner}
\end{figure}

We present the marginalized parameter constraints from our Fisher forecast for the \DESI\ BGS at $z \sim 0.3$, based on a joint analysis of $\logxicgsfull$, $\log n_{\rm gal}$, and $\log n_{\rm halo}$. We refer to this configuration as the Full Forecast. It simultaneously constrains the modified gravity parameter $\Fpar$, the cosmological parameters $\Om$ and $\sig$, the HOD parameters governing galaxy occupation and velocity bias, and the mass--observable relation parameters that describe cluster selection.

Figure~\ref{fig:full_corner} presents the marginalized $1\sigma$ and $2\sigma$ posterior ellipses for the Full Forecast, where the ellipses correspond to the 68.3\% and 95.4\% confidence regions of the joint two-parameter distributions and the diagonal panels show the marginalized one-dimensional posteriors for each parameter. We list the $1\sigma$ marginalized constraint on each parameter, obtained from the square root of the corresponding diagonal element of the inverse Fisher matrix, at the top of each diagonal panel and in the right column of table~\ref{tab:constraints}.

\begin{table}[tbp]
    \centering
    \caption{Marginalized $1\sigma$ Fisher forecast constraints for all model parameters in the Full Forecast configuration.}
    \label{tab:constraints}
    \begin{tabular}{lcc}
        \hline\hline
        Parameter & Fiducial value & $1\sigma$ constraint \\
        \hline
        $\lmcut$      & 14.00  & 0.041 \\
        $\smscat$     & 0.25   & 0.039 \\
        $\siglogM$    & 0.60   & 0.152 \\
        $\logMmin$    & 12.50  & 0.104 \\
        $\logMone$    & 13.35  & 0.068 \\
        $\alphaHOD$   & 1.00   & 0.085 \\
        $\etacen$     & 0.10   & 0.197 \\
        $\etasatlow$  & 1.00   & 0.286 \\
        $\etasathigh$ & 1.00   & 0.026 \\
        $\Fpar$       & 1.00   & 0.033 \\
        $\Om$         & 0.3152 & 0.010 \\
        $\sig$        & 0.8080 & 0.010 \\
        \hline
    \end{tabular}
\end{table}

Our joint data vector yields tight constraints on the cosmological parameters, with $\sigma(\Om) = 0.010$ and $\sigma(\sig) = 0.010$. As previously noted, we consider these forecast constraints preliminary because the derivatives with respect to $\sig$ and $\Om$ are noisy when evaluated using a single simulation volume. We constrain the phenomenological modified gravity parameter $\Fpar$ to $\sigma(\Fpar) = 0.033$, demonstrating that the \DESI\ BGS can constrain deviations from GR in the cluster infall velocity field at the 3.3\% level.

We now examine the most prominent parameter degeneracies visible in figure~\ref{fig:full_corner}, focusing on the physical mechanisms that drive each correlation. In general, the correlations are strongest between pairs of parameters that have similar structure in their derivatives in figures~\ref{fig:2D_derivatives}--\ref{fig:nd_derivatives}.

As expected, the observable mass threshold $\lmcut$ and the mass--observable scatter $\smscat$ are positively correlated. Increasing the mass threshold increases the mean cluster mass and bias and decreases the cluster number density, and increasing the log-normal scatter can compensate for all three effects. For similar reasons, the minimum mass and log-normal scatter for central galaxies, $\logMmin$ and $\siglogM$, are positively correlated. The satellite occupation parameters, $\logMone$ and $\alphaHOD$, are positively correlated because of their compensating impact on the number of satellite galaxies in cluster mass halos (recall that raising $\logMone$ \textit{lowers} the amplitude of $\langle N_{\rm sat}(M)\rangle$). The values of $\logMmin$ and $\alphaHOD$ are anti-correlated, as one would expect from their similar derivatives in figure~\ref{fig:2D_derivatives}. From these derivatives, one might expect a positive correlation between $\logMmin$ and $\logMone$, but in this case the similar impact of both parameters on $\log n_{\rm gal}$ (figure~\ref{fig:nd_derivatives}) leads to anti-correlation. The cluster selection parameters exhibit only weak correlations with the galaxy occupation parameters because their impact on $\xicgsfull$ is very different outside the FoG regime (figure~\ref{fig:2D_derivatives}).

By the same logic, we might expect an anti-correlation between $\Fpar$ and $\lmcut$, while figure~\ref{fig:full_corner} shows a mild positive correlation. We have confirmed that the expected negative correlation appears when $\smscat$ is held fixed. However, when $\smscat$ is free, the model raises $\smscat$ alongside $\lmcut$ to preserve $\log n_{\rm halo}$, introducing lower-mass halos into the sample and reducing velocities. The positive $\Fpar$--$\lmcut$ correlation is therefore largely a consequence of the stronger $\Fpar$--$\smscat$ degeneracy.

Our forecast shows a positive correlation between $\Fpar$ and $\etasathigh$, which multiplies the velocity dispersions of satellite galaxies in halos above $M = 10^{13.5}\,h^{-1}\,\Msun$. Although both parameters have a strong impact on the redshift-space distortion of $\xicgsfull$, their effects differ enough in shape that each remains individually well bounded. The derivatives of $\logxicgsfull$ with respect to $\etacen$ and $\etasatlow$ are much smaller than those for $\etasathigh$ (by roughly an order of magnitude), with the result that both parameters are poorly constrained by our observables. Our forecast shows moderate anti-correlation of $\Fpar$ with both of these parameters. The central and high-mass satellite velocity bias parameters, $\etacen$ and $\etasathigh$, are also anti-correlated with each other, since both alter the small-scale line-of-sight velocity dispersion of galaxies in cluster halos. The absence of a strong degeneracy between $\Fpar$ and any of the velocity bias parameters demonstrates that measuring clustering in redshift space does not introduce a fundamental degeneracy between modified gravity and the dynamics \textit{within} halos, despite both effects altering the velocity field.

The cosmological parameters $\sig$ and $\Om$ are both well constrained in our forecast (at the $\approx 0.01$ level) with little correlation between them. This lack of correlation shows that the information content of our data vector is quite different from that of a conventional analysis of cluster number counts versus mass, where a combination $S_8 \approx \sig \Om^{0.5}$ is constrained better than either parameter individually (e.g., \cite{White1993, Abbott2025}). The value of $\Fpar$ is moderately anti-correlated with those of $\sig$ and $\Om$, which is as expected since increasing any of these parameters leads to larger velocities in cluster infall regions, though $\sig$ and $\Om$ also affect the real-space clustering $\xicgrfull$ (see appendix~\ref{app:realspace}). We show below (figure~\ref{fig:forecast_summary}) that fixing $\sig$ and $\Om$ only leads to a small improvement in the marginalized error on $\Fpar$, and fixing $\Fpar$ leads to only a small improvement in the marginalized errors on $\sig$ and $\Om$, confirming the impression from figure~\ref{fig:full_corner} that these correlations are only moderate. We again caution that our $\sig$ and $\Om$ derivatives are noisy because they come from a single set of \AbacusSummit\ phases, and the forecast errors and error correlations could change with more accurate derivatives.

\section{Discussion}\label{sec:discussion}

\subsection{Forecast variations}
\label{sec:variations}

To better understand what drives the constraints in the Full Forecast, we consider a suite of variations in which we modify the composition of the data vector or hold specific parameters fixed. We first examine how the constraints change as we remove observables from the data vector, isolating the individual contributions of $\logxicgs$, $\log n_{\rm gal}$, and $\log n_{\rm halo}$. We then fix subsets of parameters to identify which degeneracies most limit the constraint on $\Fpar$.

\begin{figure}[tbp]
\centering
\includegraphics[width=\textwidth]{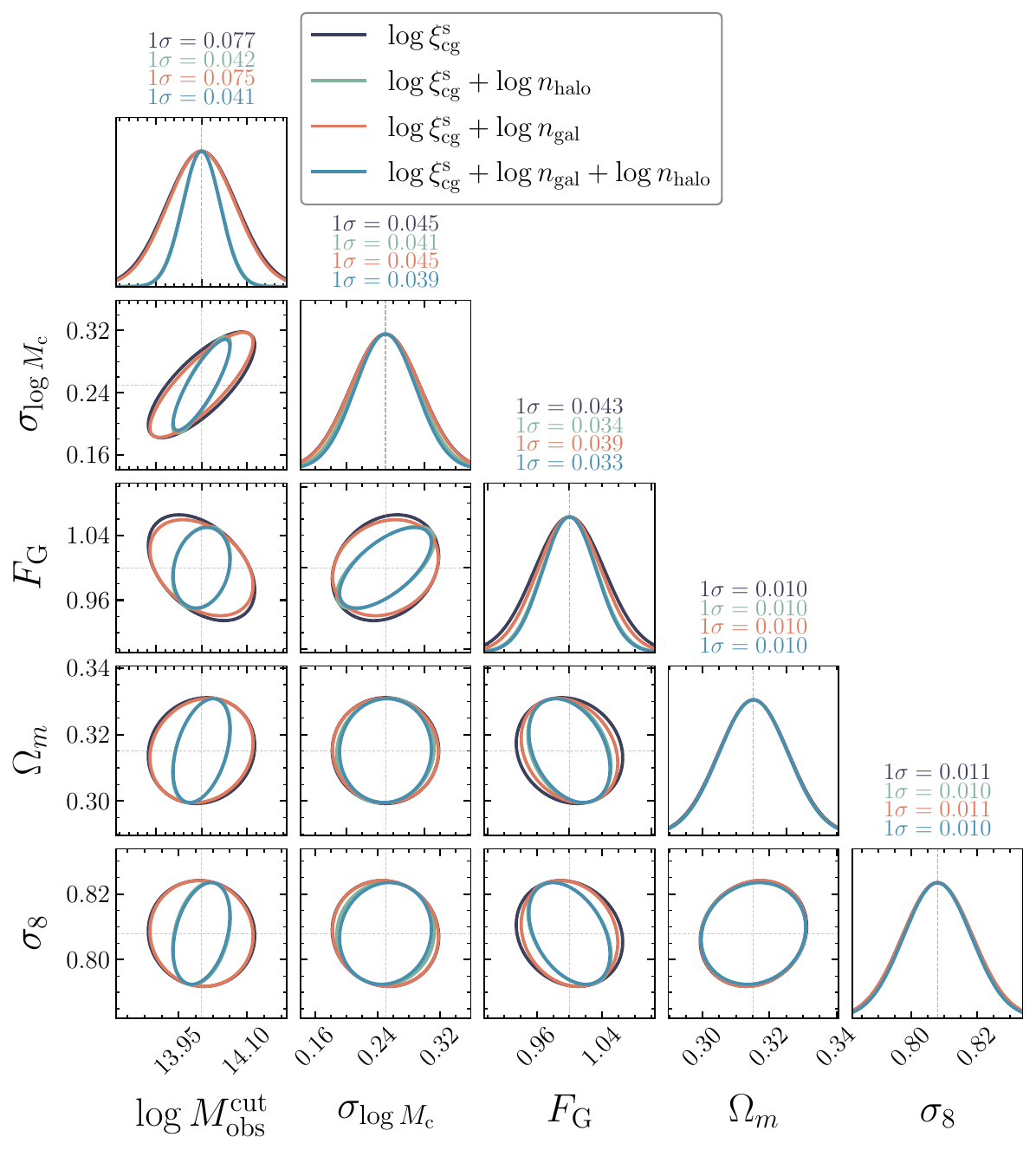}
\caption{Marginalized $1\sigma$ contours for the cluster selection, modified gravity, and cosmological parameters under four data vector configurations, $\logxicgs$ alone (navy), $\logxicgs + \log n_{\rm halo}$ (green), $\logxicgs + \log n_{\rm gal}$ (orange), and the Full Forecast combining all three observables (blue). All parameters are free in each case, and only a subset are shown for clarity. Adding $\log n_{\rm halo}$ provides the dominant improvement on $\Fpar$ by tightening the cluster selection parameters, while the cosmological parameters are largely insensitive to the inclusion of either number density.}
\label{fig:observable_variations}
\end{figure}

Figure~\ref{fig:observable_variations} presents the marginalized $1\sigma$ contours for the cluster selection, modified gravity, and cosmological parameters as we progressively add observables to the data vector, with all parameters free in each configuration. The constraint on $\Fpar$ degrades from $\sigma(\Fpar) = 0.033$ in the Full Forecast to $\sigma(\Fpar) = 0.043$ when we use only $\logxicgs$. The dominant gain from including the number densities comes from $\log n_{\rm halo}$, which alone reduces $\sigma(\Fpar)$ to $0.034$, while adding only $\log n_{\rm gal}$ reduces it to $0.039$. This hierarchy is consistent with the parameter sensitivities established in section~\ref{sec:deriv_nd}. As shown there, $\log n_{\rm halo}$ is strongly sensitive to the cluster selection parameters $\lmcut$ and $\smscat$. Providing a direct constraint on the cluster abundance tightens the constraint on $\lmcut$ by nearly a factor of two (from $0.077$ to $0.042$) and mitigates the $\Fpar$--$\smscat$ degeneracy identified in section~\ref{sec:results} as the largest single source of uncertainty on $\Fpar$. The galaxy number density, $\log n_{\rm gal}$, by contrast, is primarily sensitive to the HOD parameters, and it provides only moderate additional leverage on $\Fpar$ beyond what is already encoded in $\logxicgs$.

A notable consequence of removing $\log n_{\rm halo}$ is the reversal of the $\Fpar$--$\lmcut$ correlation visible in the upper-left off-diagonal panel of figure~\ref{fig:observable_variations}. In the Full Forecast, this correlation is positive, as discussed in section~\ref{sec:results}. When $\lmcut$ is raised, $\smscat$ must simultaneously increase to preserve the observed $\log n_{\rm halo}$, introducing lower-mass halos into the cluster sample and suppressing the effective velocity field, which requires $\Fpar$ to compensate upward. Without $\log n_{\rm halo}$ in the data vector, this compensating mechanism is no longer enforced and the correlation reverts to the physically intuitive negative sign. Raising $\lmcut$ selects more massive clusters with deeper potential wells and stronger infall signatures, which requires $\Fpar$ to decrease to reproduce the observed distortion of $\xicgsfull$.

The cosmological parameters $\Om$ and $\sig$ show little sensitivity to the inclusion of either number density, with $\sigma(\Om) = 0.010$ and $\sigma(\sig) = 0.010$--$0.011$ across all configurations. The constraining power on $\Om$ and $\sig$ derives almost entirely from the distinct signatures they imprint on $\logxicgs$.

\begin{figure}[tbp]
\centering
\includegraphics[width=\textwidth]{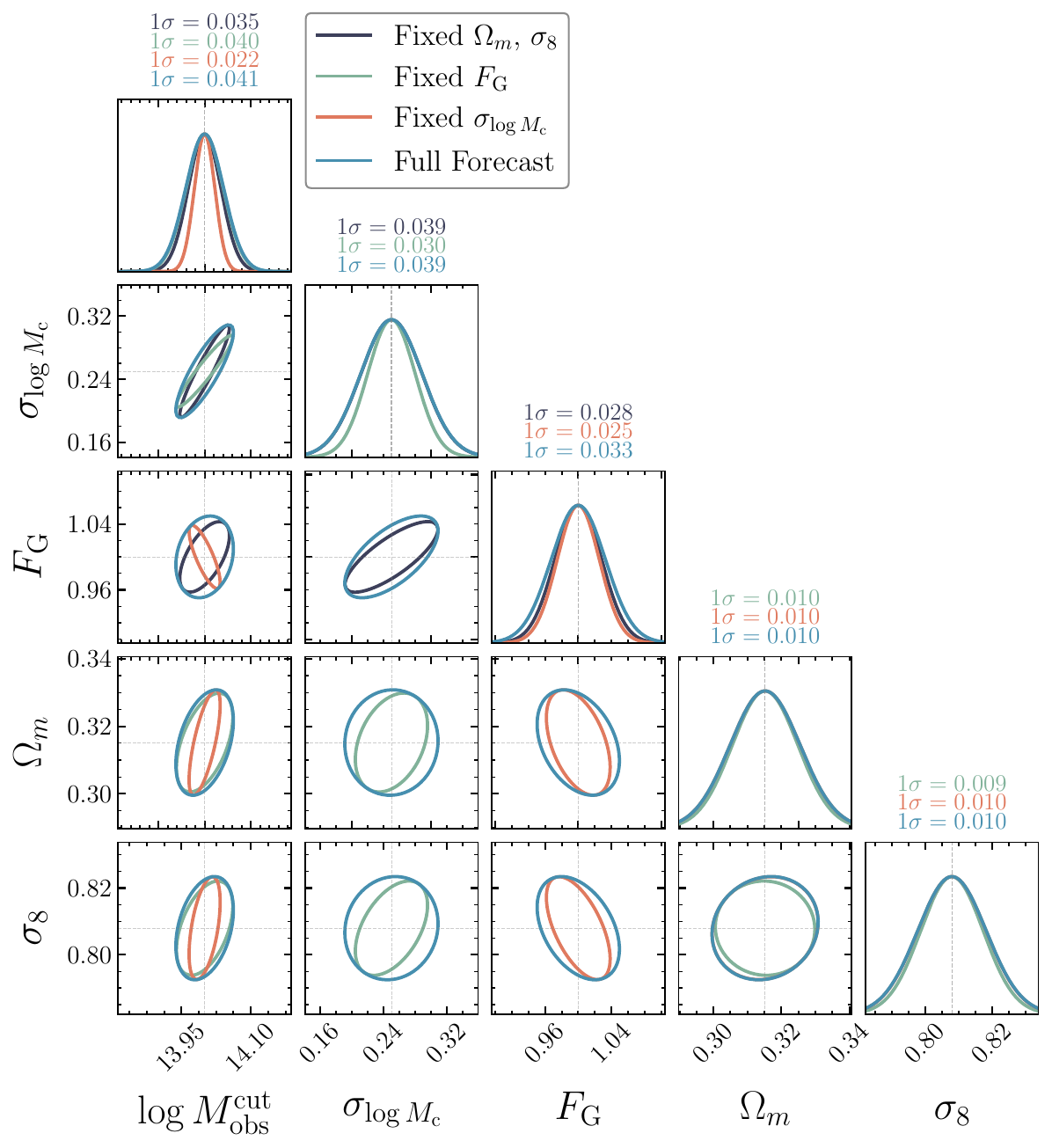}
\caption{Marginalized $1\sigma$ contours for the cluster selection, modified gravity, and cosmological parameters under four configurations, namely fixed cosmology ($\Om$ and $\sig$; navy), fixed $\Fpar$ (green), fixed $\smscat$ (orange), and the Full Forecast (blue). All other parameters remain free in each case. Contours and $1\sigma$ constraints involving a parameter are absent for the configuration in which that parameter is held fixed. Fixing $\smscat$ produces the largest improvement on $\Fpar$, from $\sigma(\Fpar) = 0.033$ to $0.025$, consistent with the $\Fpar$--$\smscat$ degeneracy being the primary limitation on the modified gravity constraint. The cosmological parameters are unaffected by fixing $\smscat$ and only modestly affected by fixing $\Fpar$, reflecting their weak degeneracy with these parameters.}
\label{fig:fixed_param_variations}
\end{figure}

To identify which parameter degeneracies most limit the constraint on $\Fpar$, figure~\ref{fig:fixed_param_variations} shows how the contours change when specific parameters are held fixed relative to the Full Forecast. We consider three cases: fixing the cosmological parameters $\Om$ and $\sig$, fixing $\Fpar$, and fixing the mass--observable scatter $\smscat$. Fixing $\smscat$ produces the largest improvement on $\Fpar$, reducing $\sigma(\Fpar)$ from $0.033$ to $0.025$, a $\sim 23\%$ improvement. This confirms that the $\Fpar$--$\smscat$ degeneracy identified in section~\ref{sec:results} is the largest limiting factor in the modified gravity constraint in the Full Forecast. Physically, $\smscat$ and $\Fpar$ are degenerate because both modulate the effective infall velocity signal. As Wu et al.~\cite{Wu2021} discuss, uncertainty in $\smscat$ also limits inferences of $\sig$ and $\Om$ from cluster weak lensing. However, with direct forward modeling of cluster selection, $\smscat$ is no longer a free parameter. Thus, while modeling the cluster selection process in DESI will introduce complications, it may improve constraints on $\Fpar$. Fixing the cosmological parameters $\Om$ and $\sig$ provides a more modest improvement in the $\Fpar$ constraint, reducing $\sigma(\Fpar)$ from $0.033$ to $0.028$ ($\sim 14\%$). Conversely, the cosmological constraints are unchanged when $\smscat$ is fixed and improve by only $6$--$9\%$ when $\Fpar$ is fixed.

\begin{figure}[tbp]
\centering
\includegraphics[width=\textwidth]{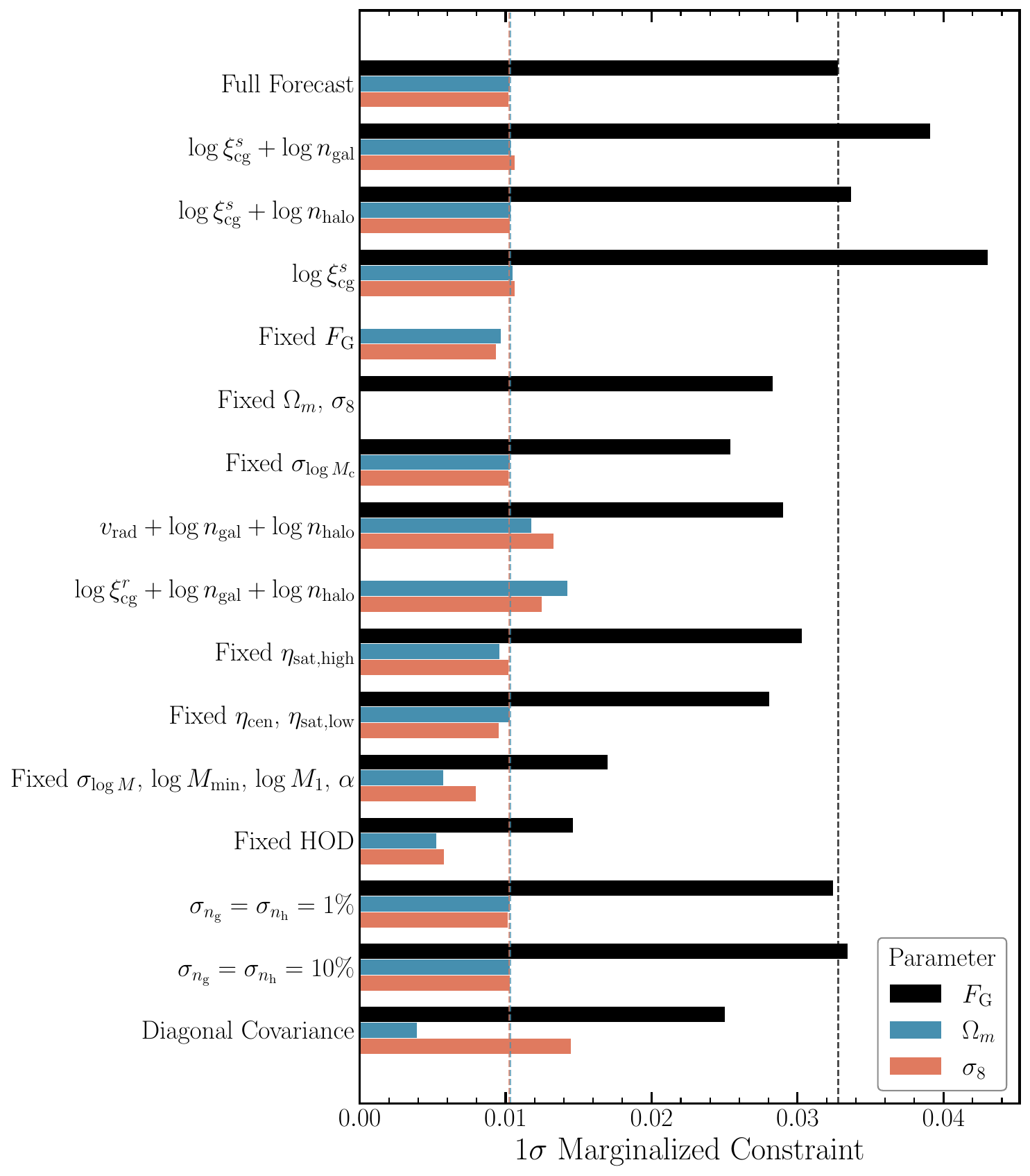}
\caption{Summary of marginalized $1\sigma$ constraints on $\Fpar$ (black), $\Om$ (blue), and $\sig$ (orange) across the full suite of forecast variations. The top row shows the fiducial Full Forecast result, and the dashed vertical lines mark these constraints for comparison against every other configuration. Rows 2--4 show the effect of removing individual observables from the Full Forecast data vector, leaving $\logxicgs$ combined with only $\log n_{\rm gal}$, only $\log n_{\rm halo}$, or neither. Rows 5--7 show the effect of fixing $\Fpar$, the cosmological parameters, or $\smscat$. Rows 8 and 9 replace the redshift-space $\logxicgs$ with the mean infall velocity profile $\vrad(r)$ and the real-space correlation function $\logxicgr$, respectively, each combined with $\log n_{\rm gal}$ and $\log n_{\rm halo}$. $\Fpar$ has no derivative with respect to $\logxicgr$ and is therefore omitted from that row. Rows 10--13 show the effect of fixing groups of HOD parameters. ``Fixed HOD'' fixes all HOD parameters, including the velocity bias parameters $\etacen$, $\etasatlow$, and $\etasathigh$. Rows 14 and 15 show the impact of setting the fractional uncertainty on $n_{\rm gal}$ and $n_{\rm halo}$ to $1\%$ and $10\%$, respectively, in place of our fiducial $5\%$ choice. Row 16 shows the result of retaining only the diagonal elements of the covariance matrix.}
\label{fig:forecast_summary}
\end{figure}

Figure~\ref{fig:forecast_summary} shows how the $\Fpar$, $\Om$, and $\sig$ errors change in many variations of our standard Full Forecast. Rows 1--7 summarize the results of figures~\ref{fig:observable_variations} and~\ref{fig:fixed_param_variations}. Rows 8--9 will be discussed in the next section. The next four rows show the impact of fixing various combinations of HOD parameters. Fixing $\etasathigh$ alone improves the $\Fpar$ error by $8\%$. While $\xicgsfull$ is much less sensitive to $\etacen$ and $\etasatlow$, these parameters are also poorly constrained, so fixing them yields a noticeable $14\%$ improvement in the $\Fpar$ error. Fixing the HOD occupation parameters $\siglogM$, $\logMmin$, $\logMone$, and $\alphaHOD$ leads to a substantial improvement in the errors on $\Fpar$, $\Om$, and $\sig$, by factors of $1.9$, $1.8$, and $1.3$, respectively, and fixing all the HOD parameters improves the errors by factors of $2.2$ ($\Fpar$), $1.9$ ($\Om$), and $1.8$ ($\sig$). In a real analysis, HOD parameters cannot be fixed \textit{a priori}, but the occupation parameters can usually be well constrained with the projected galaxy correlation function $w_{\rm{p,gg}}(\rp)$ (e.g., \cite{zehavi2011}), and it may be possible to constrain $\etacen$ and $\etasat$ effectively with the redshift-space galaxy auto-correlation $\xiggs(\rp,\rpi)$. Thus, an analysis that includes the galaxy auto-correlation in the data vector might approach the fixed-HOD errors on $\Fpar$, $\Om$, and $\sig$. However, we have not included assembly bias parameters in our HOD model, and these could degrade the constraints. We leave the investigation of extended data vectors and assembly bias to future work.

The next two rows of figure~\ref{fig:forecast_summary} show that reducing the assumed fractional errors on $\log n_{\rm gal}$ and $\log n_{\rm halo}$ from $5\%$ to $1\%$, or increasing them from $5\%$ to $10\%$, makes almost no difference to the $\Fpar$, $\Om$, and $\sig$ errors, which change by about $2\%$ at most. Omitting these number densities from the data vector entirely does degrade the $\Fpar$ constraint, as shown in figure~\ref{fig:observable_variations}, but the precision on $\Fpar$ is not sensitive to the $\log n_{\rm gal}$ and $\log n_{\rm halo}$ uncertainties over this fairly wide range. The last row of figure~\ref{fig:forecast_summary} shows a forecast that uses only the diagonal elements of the covariance matrix. This forecast shows a moderate reduction ($24\%$) in the $\Fpar$ error, a substantial factor of $2.6$ reduction in the $\Om$ error, and a $42\%$ increase in the $\sig$ error. These differences show that including off-diagonal terms in the covariance is important, but also that moderate inaccuracies in the off-diagonal correlations will not strongly affect the $\Fpar$ error.

\subsection{Constraining Power of the Infall Velocity Field}
\label{sec:infall_constraints}

\begin{figure*}[]
\centering
\includegraphics[width=\textwidth]{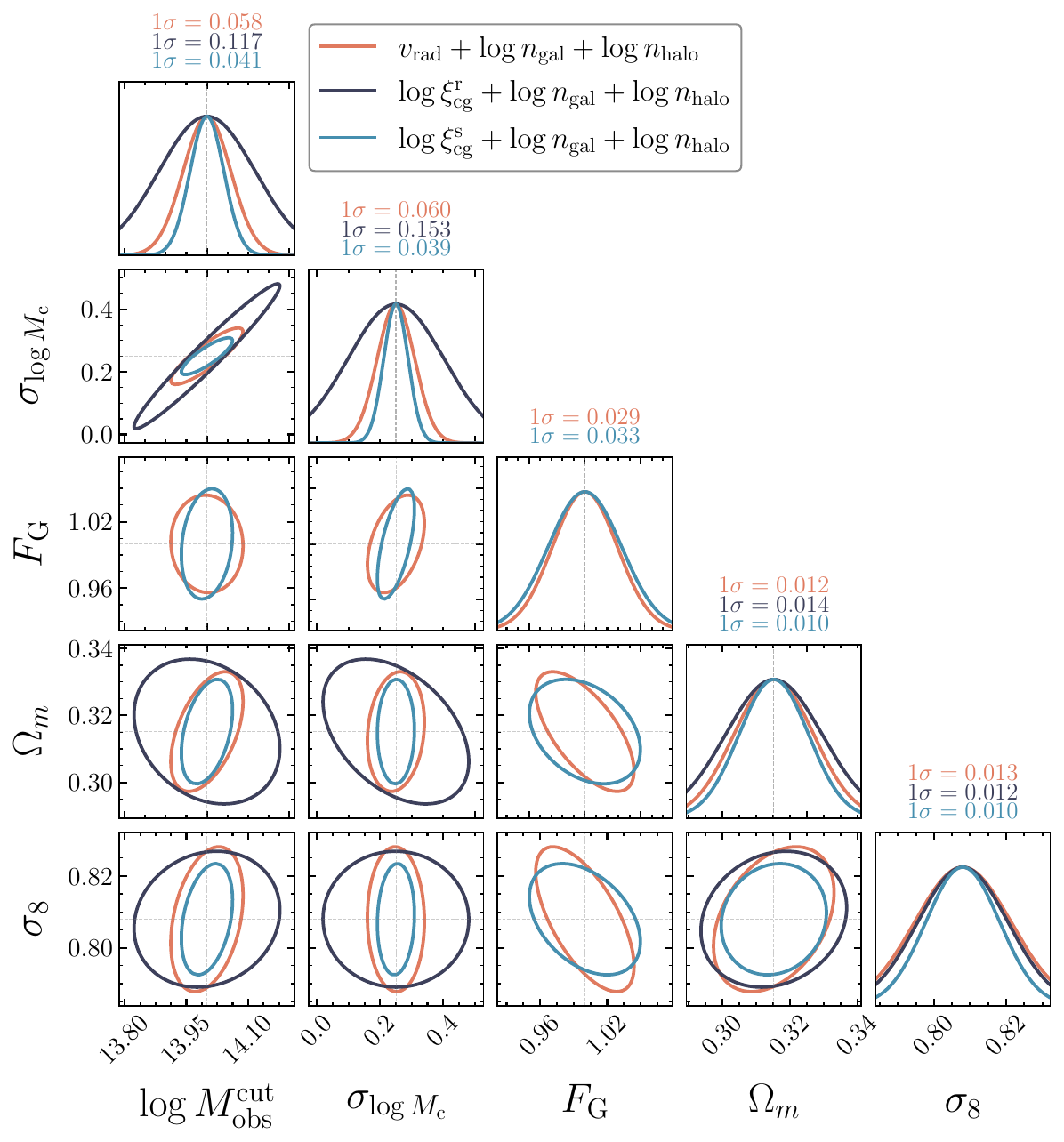}
\vspace{-5mm}
\caption{Marginalized $1\sigma$ contours for the cluster selection, modified gravity, and cosmological parameters under three data vector configurations, each combined with $\log n_{\rm gal}$ and $\log n_{\rm halo}$: the mean radial infall velocity profile $\vrad(r)$ (red), the real-space cross-correlation function $\logxicgr$ (navy), and the Full Forecast using $\logxicgs$ (blue). The velocity bias parameters and $\Fpar$ are excluded from the real-space forecast as their derivatives with respect to $\xicgr$ are zero. The constraint on $\Fpar$ from $\logxicgsfull$ is only $13\%$ weaker than from a direct measurement of $\vrad(r)$, which implies that our analysis extracts most of the information encoded in infall profiles from the redshift-space cross-correlation, even after marginalizing over HOD parameters.}
\label{fig:infall_constraints}
\end{figure*}

Zu \& Weinberg~\cite{zuw2013} presented a method to estimate infall velocity profiles $v_{\rm c}(r)$ from $\xicgsfull$, and Zu et al.~\cite{Zu2014} showed that the extracted profiles could provide a test of modified gravity models. In this paper, by contrast, we directly model $\xicgsfull$ as the primary observable. We now wish to determine whether the constraining power in $\Fpar$ actually derives from the infall velocity profile and whether the constraints are degraded relative to those that would apply if we could measure $\vrad(r)$ directly (which in practice we cannot).\footnote{Note that we define $\vrad(r)$ as the mean infall velocity, while $v_{\rm c}(r)$ from ref.~\cite{zuw2013} is a characteristic velocity that is not identical to the mean.} To address this question, we compare three data vector configurations: the Full Forecast using $\logxicgs$, a forecast using the mean radial infall velocity profile $\vrad(r)$ directly, and a forecast using the real-space cross-correlation function $\logxicgr$. In all cases we include $\log n_{\rm gal}$ and $\log n_{\rm halo}$ as observables with $5\%$ (0.022 dex) uncertainties.

Figure~\ref{fig:infall_constraints} presents the marginalized $1\sigma$ contours for the cluster selection, modified gravity, and cosmological parameters under each configuration. We marginalize over the HOD parameters but do not show them here. The constraints on cluster selection parameters are substantially better when using $\logxicgs$ instead of $\vrad(r)$, though $\vrad(r)$ outperforms the real space $\logxicgr$, showing that infall dynamics provide a key part of the information in $\logxicgs$. However, for our purposes $\lmcut$ and $\smscat$ are nuisance parameters, so the constraints on these parameters alone are not especially important.

The key result of figure~\ref{fig:infall_constraints} is that the constraint on $\Fpar$ from $\logxicgs$ is only modestly weaker than the constraint from a direct measurement of $\vrad(r)$, $\sigma(\Fpar) = 0.033$ versus $0.029$. This similarity implies that our analysis can extract most of the relevant information about $\vrad(r)$ from $\logxicgs$, even after marginalizing over HOD parameters. The real space $\logxicgr$ contains no information about $\Fpar$ because it is unaffected by peculiar velocities. In principle, $\logxicgs$ could contain additional information about $\Fpar$ from the full velocity distribution in infall regions. Still, we expect that most of the useful velocity information resides in $\vrad(r)$, especially after marginalizing over velocity bias parameters. The $13\%$ larger $\sigma(\Fpar)$ from $\logxicgs$ relative to $\vrad(r)$ likely reflects degeneracies between $\Fpar$ and the HOD and velocity bias parameters, which affect $\logxicgs$ but have little effect on $\vrad(r)$ (figure~\ref{fig:vrad}), even though $\logxicgs$ yields considerably tighter constraints on the cluster selection parameters. The real-space correlation function $\logxicgr$ already provides interestingly tight constraints on $\Om$ and $\sig$, though these improve when adding the dynamical information available in $\logxicgs$, and $\vrad(r)$ plus number densities give constraints similar to or better than those from $\logxicgr$. The summary comparison of constraints from these data vectors is shown in rows 1, 8, and 9 of figure~\ref{fig:forecast_summary}.

The real-space clustering constraints should be regarded with some caution because our HOD model for spatial clustering is somewhat restrictive; in particular, we do not include assembly bias parameters, and we assume that satellite occupation is a power law in halo mass. A more flexible model might show more degeneracy between HOD parameters and $(\sig, \Om)$, especially for $\logxicgr$, and could also degrade the $\Fpar$ constraint. Although $\vrad(r)$ itself is insensitive to the HOD (figure~\ref{fig:vrad}), marginalizing over the HOD parameters already increases $\sigma(\Fpar)$ from $\logxicgs$ by a factor of $2.2$ relative to the fixed-HOD case (figure~\ref{fig:forecast_summary}). Conversely, external HOD constraints from galaxy auto-correlations could substantially strengthen it.

\subsection{Future Work}
\label{sec:future}

The results presented here establish the statistical sensitivity achievable with the DESI BGS for testing GR-predicted infall velocities within a Fisher matrix framework. Applying this methodology to real data requires several extensions, spanning improvements to the theoretical parameterization, the forward modeling pipeline, and the observational setup.

The most immediate theoretical extension is to place the $\Fpar$ constraint in the context of specific modified gravity theories. The $f(R)$ and Galileon models investigated by Zu et al.~\cite{Zu2014} predict infall velocity enhancements of $15$--$30\%$ relative to GR at $r = 5\ h^{-1}\ \mathrm{Mpc}$, corresponding to $\Fpar \approx 1.15$--$1.30$. Our Full Forecast constraint of $\sigma(\Fpar) = 0.033$ would therefore distinguish these models from GR at $\gtrsim 4\sigma$. A natural next step is to run dedicated modified gravity simulations and fit for the best-fitting $\Fpar$ in each case, establishing a direct mapping between our phenomenological parameter and model-specific parameters such as $f_0$ in Hu--Sawicki $f(R)$ gravity \citep{HuSawicki2007}.

More generally, a uniform velocity rescaling is a simplification, as modified gravity theories generically predict enhancements that depend on scale, halo mass, and environment through their screening mechanisms. Chameleon screening (e.g., $f(R)$ gravity) and Vainshtein screening (e.g., Galileon and DGP gravity) screen via qualitatively different physical mechanisms, and these differences are expected to imprint distinct radial shapes on the infall velocity profile beyond a simple change in amplitude. We therefore plan to develop a parameterization that captures these shape differences, rather than a single scale-independent rescaling, and to characterize how this shape information propagates into the anisotropic structure of $\logxicgsfull$. This would extend our forecast beyond detecting a deviation from GR to distinguishing between specific screening mechanisms. Allowing $\Fpar$ to vary with scale or halo mass, calibrated directly against modified gravity simulations, is a first step toward this more flexible parameterization.

Applying this methodology to real DESI BGS data will require the ability to predict $\logxicgsfull$ at all observationally viable points in our parameter space, so that we can find best-fit parameters and their uncertainties. Following the approach successfully applied to galaxy--galaxy lensing and redshift-space galaxy clustering \citep{Wibking2020, Salcedo2022}, we plan to construct an emulator that interpolates predictions for $\xicgsfull$ across a grid of cosmological and HOD parameters. The derivatives computed in section~\ref{sec:sensitivity} guide us in prioritizing the parameter dimensions that most strongly affect the observable. We will need to validate the emulator against simulations and mock catalogs before applying it to the data.

A realistic application to DESI BGS will also require a more accurate model of the cluster selection function. Our current model assumes a simple log-normal scatter between observed and true halo mass with isotropic selection, but realistic cluster catalogs derived from optical richness, X-ray luminosity, or the SZ signal exhibit more complex selection functions, including projection effects and mass-dependent scatter that can introduce anisotropic biases in $\xicgsfull$ \citep{Salcedo2025, Cao2025, Lee2025, Zeng2023, Wu2022}. Developing a forward model that accurately reproduces the selection function of a DESI BGS group or cluster catalog and validating it against mock catalogs that replicate the photometric and spectroscopic selection of the BGS will be essential for unbiased inference. Our group's cluster weak lensing studies again provide a useful template, and they show that direct forward modeling of cluster selection from the galaxy population can improve cosmological constraints by effectively removing $\smscat$ as a nuisance parameter \citep{Salcedo2025, Cao2026}.

Our forecast targets a single redshift slice at $z = 0.3$, but the DESI BGS spans $0.1 < z < 0.5$, offering the possibility of measurements across multiple redshift bins. We expect modified gravity signatures to evolve with redshift, since the growth of structure and the efficiency of screening mechanisms depend on the cosmic epoch, so splitting the BGS into multiple bins would provide an additional discriminant between models and a consistency check against GR.

As noted in section~\ref{sec:deriv_cosmology}, the derivatives with respect to $\Om$ and $\sig$ come from single-paired simulation phases and are consequently noisier than the HOD and mass--observable derivatives, introducing uncertainty into our cosmological forecasts. We hope that constructing an emulator that interpolates across a range of cosmological parameters will reduce this noise and improve the reliability of the $\Om$ and $\sig$ constraints. We also need to test the possible impact of a more flexible HOD parameterization, especially the impact of assembly bias or a more general $\langle N_{\rm sat}(M) \rangle$. As discussed in section~\ref{sec:infall_constraints}, a more flexible HOD could weaken our constraints on $\sig$ and $\Om$, and it could also degrade the $\Fpar$ constraint, which is already a factor of $2.2$ weaker than in the fixed-HOD case. Conversely, as discussed in section~\ref{sec:variations}, adding the galaxy projected and redshift-space autocorrelation functions to the data vector may tighten HOD constraints and, in turn, tighten constraints on $\Fpar$, $\Om$, and $\sig$.
\section{Conclusions}\label{sec:conclusions}

We have presented a Fisher matrix forecast for the precision with which the DESI BGS can constrain departures from GR in the cluster infall velocity field, using the redshift-space galaxy--cluster cross-correlation function $\xicgsfull$ as the primary observable. Our analysis simultaneously marginalizes over the HOD parameters governing the galaxy--halo connection, the mass--observable relation parameters describing the cluster selection, and the cosmological parameters $\Om$ and $\sig$, providing a realistic assessment of the statistical sensitivity achievable in the presence of these astrophysical uncertainties. The main conclusions of this work are as follows.

\begin{itemize}

\item The \DESI\ BGS can constrain the phenomenological modified gravity parameter $\Fpar$ to $\sigma(\Fpar) = 0.033$ at $z \sim 0.3$, enabling the detection of deviations from GR in the cluster infall velocity field at the 3.3\% level. Based on the results of Zu et al.~\cite{Zu2014}, this sensitivity is sufficient to distinguish $f(R)$ and Galileon gravity models from GR at $\gtrsim 4\sigma$, representing a substantial improvement over the $\sim 10$--$30\%$ uncertainties that limited earlier SDSS analyses.

\item The modified gravity parameter $\Fpar$ is only moderately correlated with the HOD parameters governing galaxy occupation and velocity bias, demonstrating that we can measure the cluster infall velocity signal without introducing a fundamental degeneracy with the galaxy--halo connection. The strongest limitation on the $\Fpar$ constraint is its degeneracy with the mass--observable scatter $\smscat$, which is partially broken by the independent constraint on the cluster abundance. Fixing $\smscat$ improves the constraint on $\Fpar$ by $\sim 23\%$.

\item The redshift-space cross-correlation function $\logxicgs$ encodes both the infall velocity information that drives the $\Fpar$ constraint and the spatial clustering information that constrains the nuisance parameters. Replacing $\logxicgs$ with a direct measurement of the mean radial infall velocity profile $\vrad(r)$ gives only a modestly tighter constraint on $\Fpar$ ($0.029$ versus $0.033$) but substantially weaker constraints on the cluster selection parameters, so modeling $\logxicgs$ directly recovers most of the infall information.

\item We forecast constraints on the cosmological parameters $\Om$ and $\sig$ of $\sigma(\Om) = 0.010$ and $\sigma(\sig) = 0.010$, respectively, with their constraining power deriving almost entirely from the distinct radial and anisotropic signatures they imprint on $\logxicgs$ rather than from the number density observables. The $\Fpar$ parameter is only moderately anti-correlated with $\Om$ and $\sig$, establishing that we can separate the modified gravity and cosmological constraints without significant mutual degradation. Fixing $\Om$ and $\sig$ allows a slightly improved constraint on $\Fpar$, and fixing $\Fpar=1$ allows slightly improved constraints on $\Om$ and $\sig$. 

\item The sensitivity demonstrated here motivates a program of follow-up work, including calibration of $\Fpar$ against dedicated modified gravity simulations to connect our phenomenological parameter to specific theories such as $f(R)$ and Galileon gravity, construction of an emulator for fast likelihood evaluation across the full parameter space, development of a more realistic cluster selection forward model, and measurements across multiple DESI BGS redshift bins to probe the redshift evolution of gravitational growth.

\end{itemize}

These results establish the DESI BGS as a powerful testbed for gravitational physics on quasi-linear scales, and position $\xicgsfull$ as a compelling complement to existing probes of modified gravity. The cluster infall regime probed by $\xicgsfull$ occupies a physically distinct region in the landscape of gravity tests, as it lies beyond the screened, virialized cluster interior where fifth forces are suppressed, yet within scales where coherent infall flows are strong and well resolved by spectroscopic surveys. This is the environment where departures from GR are predicted to be most pronounced, and where the gravitational potentials governing light deflection and matter acceleration may diverge in ways that cause lensing and kinematic mass estimates to be mutually inconsistent \citep{Zu2014, Reyes2010}. Combining $\xicgsfull$ measurements with weak gravitational lensing around the same cluster sample would enable a novel gravitational slip test in this regime, probing the ratio of the two scalar metric potentials at few-percent precision in a physically distinct environment from existing $E_{\rm{G}}$ measurements \citep{Rauhut2025}. Looking further ahead, the Nancy Grace Roman Space Telescope will provide weak lensing measurements of unprecedented depth and source density over wide fields, enabling a direct cross-correlation between the kinematic infall signal measured in DESI BGS and the lensing mass profiles from Roman. Such a combination would constitute a powerful simultaneous test of whether the gravitational potentials governing light and matter are equal, as GR demands, offering one of the most direct probes of modified gravity on cosmological scales achievable in the near future.

\appendix
\section{Real-Space Cross-Correlation Function}
\label{app:realspace}

\begin{figure*}[]
\centering
\includegraphics[width = \textwidth]{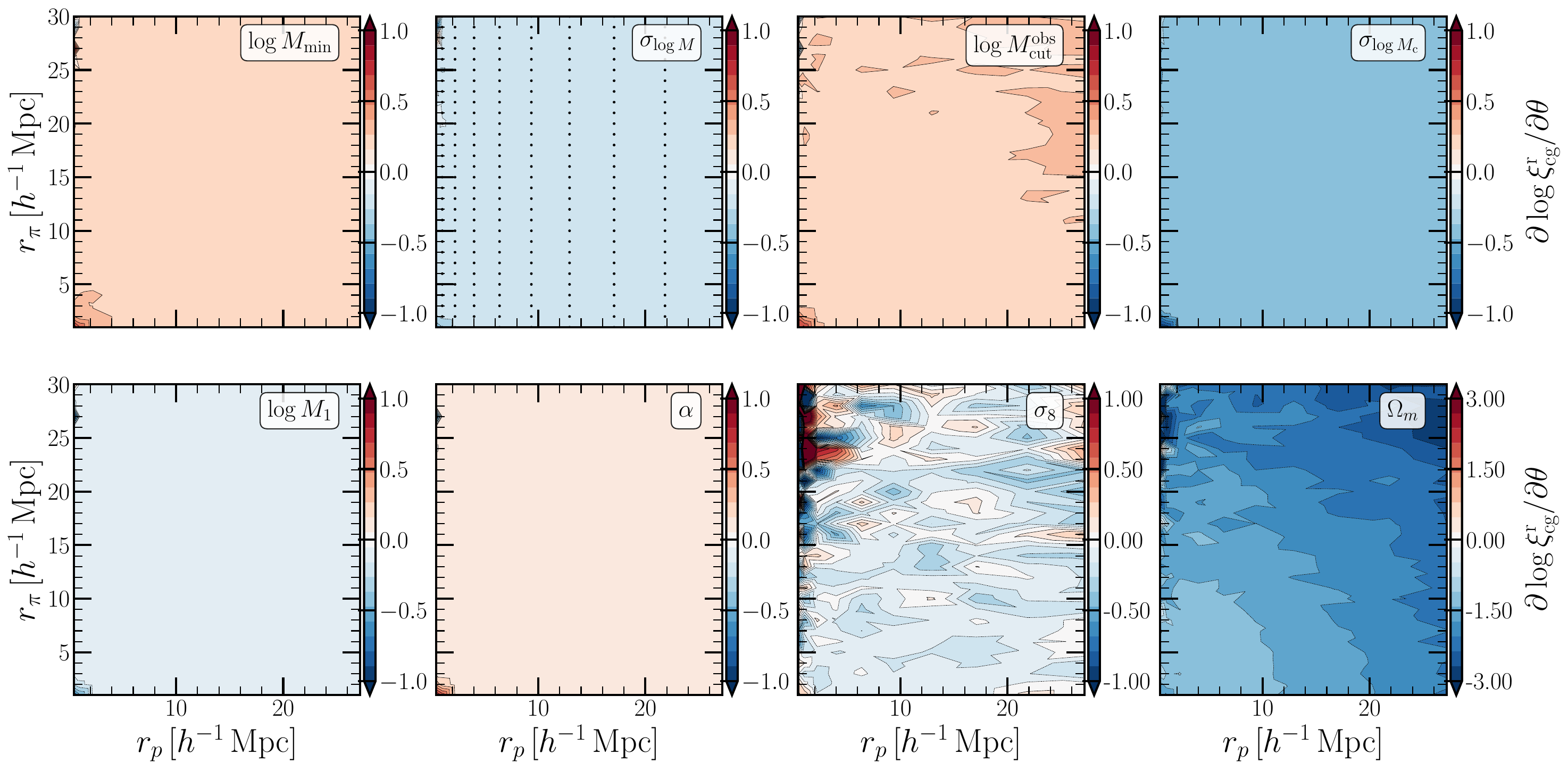}
\vspace{-5mm}
\caption{Real-space counterparts of the redshift-space derivatives shown in figures~\ref{fig:2D_derivatives} and~\ref{fig:cosmo_derivatives}, displaying $\partial \log \xi_{\rm cg}^{\rm r} / \partial \theta$ throughout the $(\rp, \rpi)$ plane. The velocity bias parameters ($\etacen$, $\etasatlow$, $\etasathigh$) and the modified gravity parameter $\Fpar$ are omitted because their real-space derivatives are zero. As in figure~\ref{fig:2D_derivatives}, red and blue hues indicate positive and negative derivatives, respectively. Isolated features at $\rp < 1\,h^{-1}\,\mathrm{Mpc}$ and $\rpi \gtrsim 25\,h^{-1}\,\mathrm{Mpc}$ are noise from sparsely populated bins. Note that we use a different color scale for the $\Om$ derivative, as figure~\ref{fig:cosmo_derivatives}.}
\label{fig:realspace_derivatives}
\end{figure*}

We compute the real-space derivatives $\partial \log \xicgrfull / \partial \theta$ following the same procedure used for the redshift-space derivatives in section~\ref{sec:sensitivity}, applying the centered finite-difference approximation of equation~(\ref{eq:deriv}) to $\logxicgrfull$ in place of $\logxicgsfull$. As in the redshift-space case, we average the HOD and mass--observable parameter derivatives over the 25 \textit{Planck} fiducial cosmology phases of \AbacusSummit, and compute the $\Om$ and $\sig$ derivatives from the paired simulation boxes listed in table~\ref{tab:cosmo}. Figure~\ref{fig:realspace_derivatives} shows the resulting real-space derivatives. The derivatives with respect to the velocity bias parameters and $\Fpar$ vanish in real space and are not shown. We discuss the interpretation of these derivatives and their relationship to the corresponding redshift-space derivatives in section~\ref{sec:sensitivity}.

\section{Spherical Model for the $\Fpar$ Derivative}
\label{app:model}

\newcommand{\vq}{{\bf q}}
\newcommand{\vx}{{\bf x}}
\newcommand{\vr}{{\bf r}}
\newcommand{\rhoint}{{\rho_{\rm int}}}

To build intuition for the redshift-space distortion of $\xicgsfull$, we consider an approximate description based on the infall region of a single, spherically symmetric cluster, extending the model presented by Kaiser~\cite{Kaiser1987}. As a spherical perturbation evolves, a shell of matter with Lagrangian radius $q$ in the uniform initial conditions moves inward to a radius $r$, and the cumulative overdensity profile is
\begin{equation}
    \rhoint(r) = \left[\frac{q(r)}{r}\right]^3\,,
    \label{eqn:rhobar}
\end{equation}
where $\rhoint(r)$ is the mean interior overdensity in units of the cosmic mean density $\bar{\rho}$. For a spherical perturbation in an $\Omega_m=1$ universe, the overdensity $1+\delta$ can be written in terms of a timelike variable $\eta_P$ as
\begin{equation}
    1+\delta = \left(\frac{q}{r}\right)^3 = \frac{9}{2} (\eta_P-\sin\eta_P)^2(1-\cos\eta_P)^{-3}\,,
    \label{eqn:sphrho}
\end{equation}
with collapse occurring at $\eta_P=2\pi$ (\cite{Gunn1972}; for a textbook treatment see, e.g., ref.~\cite{Padmanabhan1993}). One can model a spherical overdensity with profile $\rhoint(r)=1+\delta(r)$ as a nested set of perturbations with $\eta_P$ decreasing outward as implied by equation~\eqref{eqn:sphrho}. The overdensity $\rho_{\rm vir} = 180$, commonly taken to represent the overdensity within the virial radius, corresponds to $\eta_P \approx 4.8$. The radial peculiar velocity can be written
\begin{equation}
    v_{\rm rad,sph} = Hr \left[1-\frac{3}{2}\sin\eta_P(\eta_P-\sin\eta_P)(1-\cos\eta_P)^{-2}\right]\,,
    \label{eqn:vradsph}
\end{equation}
where negative values denote radial infall. For an $\Omega_m < 1$ universe, we make the approximation of multiplying the non-linear velocity $v_{\rm rad,sph}$ by the linear growth rate factor $f(\Omega_m)$, and we further multiply by our modified gravity parameter $F_G$ to obtain
\begin{align}
    v_{\rm rad} &= Hr \left[1-\frac{3}{2}\sin\eta_P(\eta_P-\sin\eta_P)(1-\cos\eta_P)^{-2}\right]f(\Omega_m)\,F_G
    \label{eqn:vrad1}\\
      &= Hr\,Q(\rhoint)\left(1-\rhoint^{1/3}\right)f(\Omega_m)\,F_G\,.
      \label{eqn:vrad2}
\end{align}
We find the dimensionless factor $Q(\rhoint)$ by numerically solving for $\eta_P(\rhoint)$ using equation~\eqref{eqn:sphrho}. The value of $Q(\rhoint)$ rises from 1.0 at $\rhoint=1$ to about 2.4 at $\rhoint=180$, with a roughly linear dependence on $\log\rhoint$. For $Q(\rhoint)=F_G=1$, equation~\eqref{eqn:vrad2} is equivalent to the peculiar velocity implied by the Zeldovich approximation~\cite{Zeldovich1970},
\begin{equation}
    \mathbf{v}_{\rm Zel} = H(\mathbf{r}-\mathbf{q})\,f(\Omega_m)\,,
\end{equation}
where $\mathbf{r}$ is the final position of a particle with Lagrangian position $\mathbf{q}$.

\begin{figure}[t]
\centering
\includegraphics[width=0.8\textwidth]{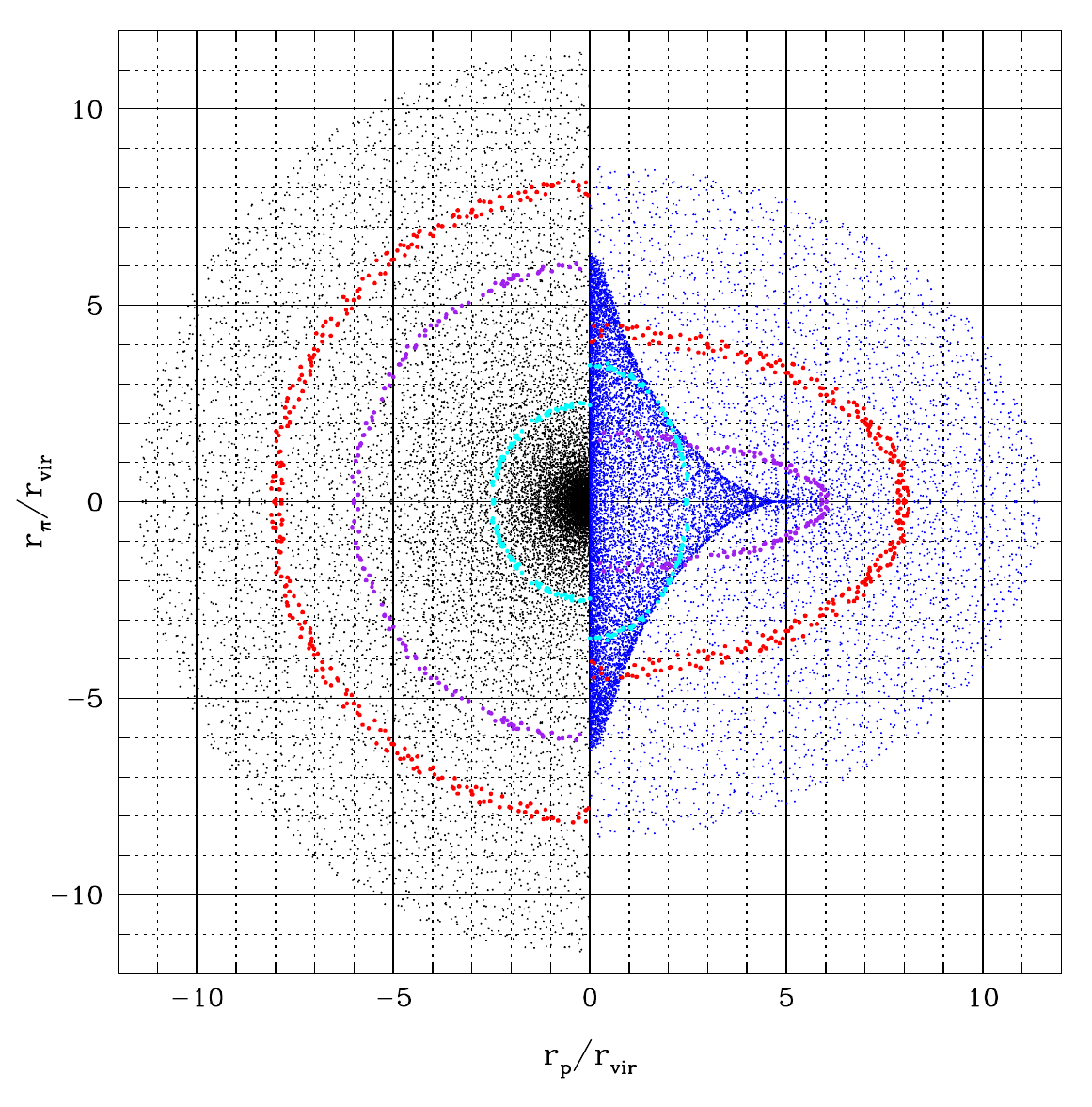}
\caption{Slice through a particle realization of a spherical overdensity in real space (left) and redshift space (right), with the line of sight in the vertical ($r_\pi$) direction. The overdensity has a cumulative profile $\rhoint(r) = 1+\rho_{\rm vir}(r_{\rm vir}/r)^2$ in units of the cosmic mean density, with $\rho_{\rm vir}=180$, and radii are scaled to $r_{\rm vir}$. The thickness of the slice is $0.5\,r_{\rm vir}$, and all particles in the slice with $r/r_{\rm vir}<11.5$ are plotted, in black on the left and blue on the right. Larger red, purple, and cyan dots show particles with $r/r_{\rm vir} = 8 \pm 0.2$, $6\pm 0.1$, or $2.5\pm 0.05$, respectively. The two larger shells are compressed along the line of sight, and the innermost shell is inverted, with infall velocities large enough that particles on the near side of the cluster appear on the far side, and vice versa. At $r_p/r_{\rm vir}<5$, redshift-space caustics create a characteristic ``trumpet'' pattern~\cite{Kaiser1987}.}
\label{fig:rhocg_particles}
\end{figure}

For a particle with real-space position $(r_p,r_\pi)$ in the projected and line-of-sight directions, the redshift-space position is $s_p = r_p$ and
\begin{equation}
    s_\pi = r_\pi + \frac{v_{\rm rad}}{H}\,\mu\,,
    \label{eqn:spi}
\end{equation}
where $\mu = r_\pi/r$ is the cosine of the angle between $\mathbf{r}$ and the line of sight. Figure~\ref{fig:rhocg_particles} shows the real-space (left) and redshift-space (right) positions of particles in a spherical overdensity with profile
\begin{equation}
    \rhoint = 1 + \rho_{\rm vir}(r_{\rm vir} / r)^2
\end{equation}
with $\rho_{\rm vir} = 180$. The value of $r_{\rm vir}$ defines the characteristic scale of the model, and a comoving value of $r_{\rm vir}=1\hmpc$ corresponds to a mass $M_{\rm vir} \approx 10^{14}\,h^{-1} M_\odot$. Particles are initially distributed in a grid of Lagrangian cells of size $0.333\,r_{\rm vir}$, with random positions within each cell to suppress artifacts of a regular grid. They are displaced radially inward to produce the desired density profile, and radial velocities are computed from equation~\eqref{eqn:vrad2} with $F_G=1$ and $f(\Omega_m)=(0.485)^{0.55}=0.67$, corresponding to the value of $\Omega_m$ at $z=0.3$ in a flat $\Lambda$CDM model with $\Omega_m(z=0)=0.3$. We set the maximum value of $Q(\rhoint)$ to 2.4, the value at $r=r_{\rm vir}$. We plot particles with $r/r_{\rm vir}<11.5$, corresponding to $\rhoint > 2.36$.

Particles colored red, purple, and cyan are those with radius $r \approx 8$, 6, or $2.5\,r_{\rm vir}$, respectively. In redshift space, the two outer shells are compressed along the line of sight, with compression at $r_p=0$ of about a factor of two for $r=8\,r_{\rm vir}$ and about a factor of three for $r=6\,r_{\rm vir}$. For $r=2.5\,r_{\rm vir}$, peculiar velocities are large enough that particles on the far side of the cluster are shifted to appear on the near side and vice versa. The full particle distribution at $r_p < 5\,r_{\rm vir}$ shows the characteristic ``trumpet'' pattern identified by Kaiser~\cite{Kaiser1987} (see his figure~5), bounded by caustics at which $ds_\pi/dr_\pi = 0$. These caustics formally have infinite density, but also infinitesimal volume and thus finite mass. At $r_p=0$ the trumpet extends to $r_\pi/r_{\rm vir} \approx \pm 6.2$. For our adopted $\rho_{\rm vir}$ and $\Omega_m$, a 1D velocity dispersion $\sigma_{\rm 1D} = (GM_{\rm vir}/3r_{\rm vir})^{1/2}$ corresponds to $\sigma_{\rm 1D}/H \approx 3.8\,r_{\rm vir}$, so the velocity elongation from infall at $r_p \approx 0$ is similar in length to the FoG from the virial velocity dispersion, though it truncates more sharply.

\begin{figure}[t]
\centering
\includegraphics[width=0.9\textwidth]{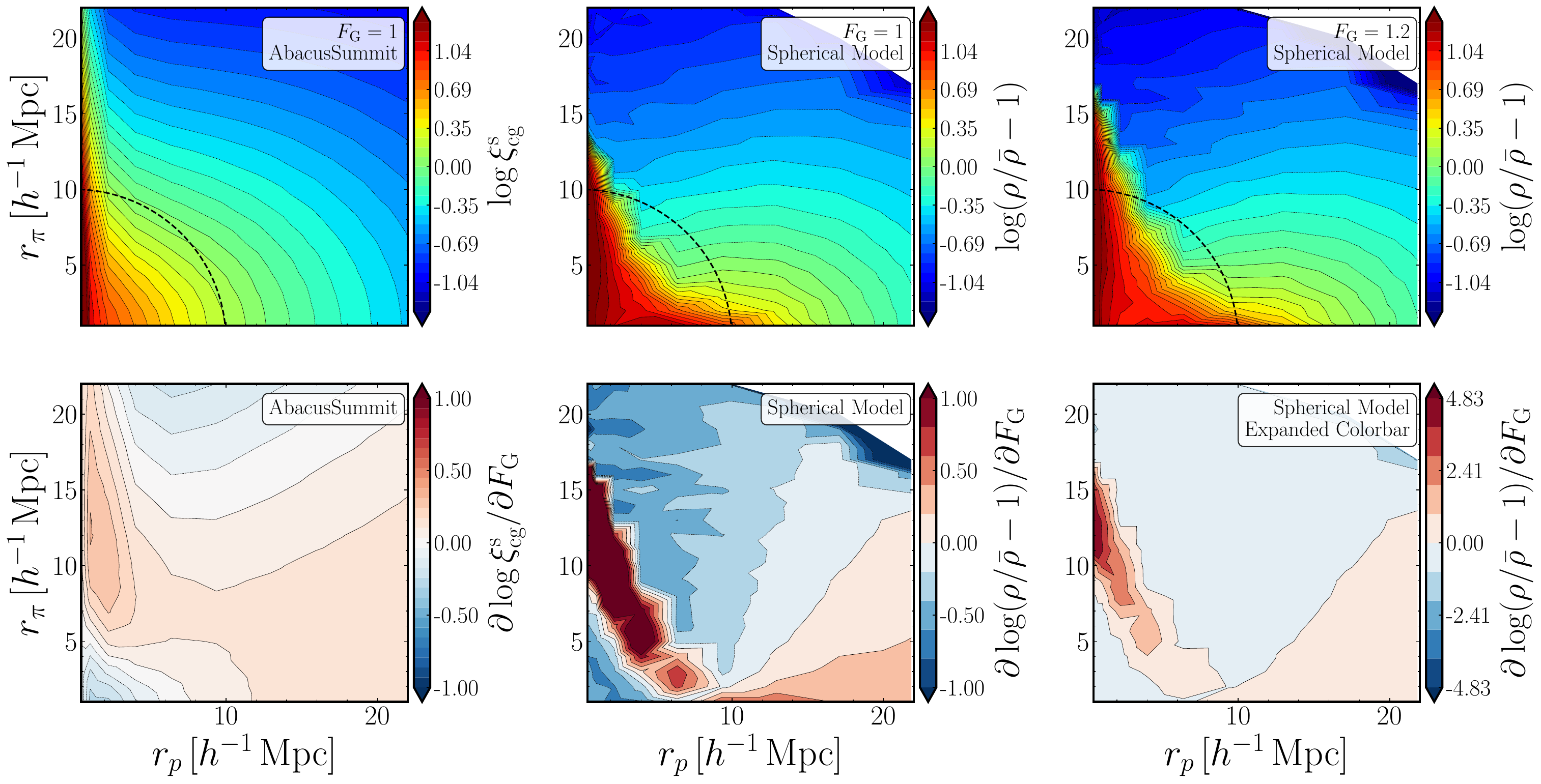}
\caption{\textbf{Top row:} contours of $\log\xicgsfull$ (repeated from figure~\ref{fig:xicgs_rprpi}) compared to contours of $\log[\rho(r_p,r_\pi)/\bar\rho-1]$ computed for the spherical overdensity model with $F_G=1.0$ (middle) and $F_G=1.2$ (right), as described in the text. \textbf{Bottom row:} derivatives of $\log\xicgsfull$ with respect to $F_G$ (repeated from figure~\ref{fig:2D_derivatives}) compared to derivatives of $\log[\rho(r_p,r_\pi)/\bar\rho-1]$ computed from the spherical model. The white regions in the upper right corners of the model panels lie beyond the largest separation computed for the spherical model and contain no model predictions. The left and middle panels use the same color scale, while the right panel adopts a broader color scale to reveal detail in the derivative structure.}
\label{fig:sphcontours}
\end{figure}

To create a model for the $F_G$ derivative, we average the overdensity profiles for $r_{\rm vir}=1.0$, 1.2, 1.4, 1.6, 1.8, and $2.0\hmpc$, since the density caustics create unrealistically sharp features if we consider a single value of $r_{\rm vir}$ (equivalent to a single cluster mass). We adopt a finer Lagrangian grid spacing of $0.16\hmpc$ to suppress particle noise. In figure~\ref{fig:sphcontours}, the top row compares the contours of $\xicgsfull$ from figure~\ref{fig:xicgs_rprpi} to the redshift-space contours of $\rho(r_p,r_\pi)/\bar\rho-1$ from the spherical model for $F_G=1.0$ and 1.2, where $\rho(r_p,r_\pi)/\bar\rho$ is the \textit{local} overdensity (not the mean interior overdensity), and we subtract one to get the quantity most directly comparable to $\xicgsfull$. We compute densities on the same grid of points used for our $\log\xicgs$ contours and derivative calculations (see the $\siglogM$ panel of figure~\ref{fig:realspace_derivatives}). The density contours show large-scale flattening and a high-density ridge along the caustic locus, as seen in the particle slice, with a transition between these features at $r_p \approx 5$--$10\hmpc$ and $r_\pi < 5\hmpc$. At $r_p < 2\hmpc$ the density maximum is displaced from $r_\pi=0$ because the infall velocity of particles within $r_{\rm vir}$ displaces them to large $r_\pi$. Virialization within $r_{\rm vir}$ would change the line-of-sight velocity distribution to be roughly Gaussian, altering the appearance of the FoG feature. The density contours of the spherical model qualitatively resemble the contours of $\xicgsfull$, but the $\xicgs$ contours show a more elongated FoG feature, less large-scale flattening, and less crowding of contours in the region of density caustics. Smooth radial infall is the only source of redshift-space distortion in the spherical model, and it is unsurprising that its features are diluted in the more complex 3D structure of the $N$-body simulations.

The bottom row of figure~\ref{fig:sphcontours} compares the $F_G$ derivatives of $\log[\rho(r_p,r_\pi)/\bar\rho-1]$ to those of $\log\xicgs(r_p,r_\pi)$. Adding a virial FoG to the spherical model would not change these derivatives because the velocity dispersions within halos are not affected by $F_G$.\footnote{Strictly speaking, the \textit{linear} derivatives would not be affected, but the logarithmic derivatives could change slightly.} In the spherical model, the strongest effect of increasing $F_G$ is to shift caustics outward in $r_\pi$ at a given $r_p$, as one can see by comparing the $F_G=1.0$ and $F_G=1.2$ contours in the top row. As a result, the density derivatives are large and positive along the diagonal ridge that traces the caustics. The derivatives are negative underneath this ridge, in the triangle at small $(r_p,r_\pi)$, because larger peculiar velocities move particles out of this region towards larger $r_\pi$. The increased flattening of large-scale density contours with larger $F_G$ leads to positive derivatives at small $r_\pi$ and $r_p=5$--$15\hmpc$, but the $F_G$ dependence is small at larger $r_\pi$ because the peculiar velocities are small at these distances.

The spherical model explains several of the key qualitative features of the $\xicgs$ derivatives, though it does not reproduce them quantitatively. Most importantly, it explains why the derivative is largest and positive at small $r_p$ and $r_\pi = 10$--$15\hmpc$, because larger $F_G$ stretches the infall features (the ``Kaiser trumpet'') along the line of sight. In similar fashion, it explains the triangle of negative derivatives at small $(r_p,r_\pi)$, and large-scale flattening explains the positive derivatives at large $r_p$ and small $r_\pi$. The spherical model is idealized, and it predicts sharper features in the derivatives than the full $N$-body distributions. It also does not explain the structure at large $(r_p,r_\pi)$, where our simple density profile may provide a poor match to the correlation function. These regions carry little weight in our forecasts because the measurement errors are largest there.

\acknowledgments

We thank Tessa Baker, Shaun Cole, and Peder Norberg for helpful discussions that shaped several aspects of this work. ANS was supported by Department of Energy (DOE) grants DE-SC0009913 and DE-SC0025993. CL was supported by an NSF Postdoctoral Fellowship under award 2502789. This work used the \textsc{AbacusSummit} suite of large, high-accuracy cosmological $N$-body simulations, produced with the \textsc{Abacus} $N$-body code~\cite{garrison21, maksimova21}. \textsc{AbacusSummit} was run on the Summit supercomputer at the Oak Ridge Leadership Computing Facility through projects AST135 and AST145. Data analysis and visualization were performed with \texttt{Corrfunc}~\cite{Sinha2020, manodeep_sinha_2020_3634195}, \texttt{NumPy}~\cite{harris2020array}, \texttt{SciPy}~\cite{2020SciPy-NMeth}, \texttt{Matplotlib}~\cite{Hunter:2007}, and SM\footnote{\url{https://www.astro.princeton.edu/~rhl/sm/}}. The simulations were analyzed in part using computational resources from the Ohio Supercomputer Center~\cite{OhioSupercomputerCenter1987}. This research has used NASA's Astrophysics Data System. This work is dedicated to the loving memory of  Czes\l{}aw Kroenke. Niech spoczywa w pokoju.


\bibliographystyle{JHEP}
\bibliography{clusters}


\end{document}